\documentclass[prd,twocolumn,showpacs,amsmath,amssymb,nofootinbib,superscriptaddress]{revtex4}
\usepackage{graphicx}
\usepackage{epsfig}
\usepackage{bm}
\usepackage{amsfonts}
\usepackage[latin1]{inputenc}
\usepackage{amssymb}
\usepackage{color}
\usepackage{float}
\usepackage{amsmath}
\usepackage{dcolumn}
\usepackage{hyperref}

\renewcommand{\]}{\right]}

\def\nn{\nonumber}

\def\be{\begin{equation}}
\def\ee{\end{equation}}
\def\bea{\begin{eqnarray}}
\def\eea{\end{eqnarray}}
\def\eqi{\begin{equation}}
\def\eqf{\end{equation}}
\def\eqia{\begin{eqnarray}}
\def\eqfa{\end{eqnarray}}
\def\lcdm{$\Lambda$CDM }
\newcommand{\fs}{{\rm{\it f\sigma_8}}}
\begin{document}

\title{Viable $f(T)$ models  are practically indistinguishable from $\Lambda$CDM}

\author{S. Nesseris}\email{savvas.nesseris@uam.es}
\affiliation{Instituto de F\'isica Te\'orica UAM-CSIC, Universidad Auton\'oma de Madrid, Cantoblanco, 28049 Madrid, Spain}

\author{S. Basilakos}\email{svasil@academyofathens.gr}
\affiliation{Academy of Athens, Research Center for Astronomy and
Applied Mathematics, Soranou Efesiou 4, 11527, Athens, Greece}

\author{E. N. Saridakis}
\email{Emmanuel\_Saridakis@baylor.edu}
\affiliation{Physics Division, National Technical University of Athens,
15780 Zografou Campus,  Athens, Greece}
\affiliation{Instituto de F\'{\i}sica, Pontificia Universidad de Cat\'olica
de Valpara\'{\i}so, Casilla 4950, Valpara\'{\i}so, Chile}

\author{L. Perivolaropoulos}\email{leandros@uoi.gr}
\affiliation{Department of Physics, University of Ioannina, 45110 Ioannina, Greece}


\begin{abstract}
We investigate the cosmological predictions of several $f(T)$ models, with up to two
parameters, at both the background and the perturbation levels. Using current cosmological
observations (geometric supernovae type Ia, cosmic microwave background and baryonic acoustic
oscillation and dynamical growth data) we impose constraints on the distortion parameter, which quantifies the deviation of these models from the concordance $\Lambda$ cosmology at the background level. In addition we constrain the growth index $\gamma$ predicted in the context of these models using the latest perturbation growth data in the context of three parametrizations for $\gamma$. The evolution of the best fit effective Newton
constant, which incorporates the $f(T)$-gravity effects, is also obtained along with the
corresponding $1\sigma$ error regions. We show that all the viable parameter sectors of
the $f(T)$ gravity models considered practically reduce these models to $\Lambda$CDM.
Thus, the degrees of freedom that open up to $\Lambda$CDM in the context of  $f(T)$ gravity
models are not utilized by the cosmological data leading to an overall disfavor of these models.
\end{abstract}

\maketitle


\section{Introduction}
The \lcdm model is currently the simplest model consistent with practically all cosmological observations. It assumes homogeneity and isotropy on large cosmological scales and the presence of a cosmological constant $\Lambda$ in the context of general relativity. Despite of its simplicity and its overall consistency with observations, \lcdm has two weak points:
\begin{enumerate}
\item
It requires a theoretically unnatural and fine-tuned value for $\Lambda$.
\item
It is marginally consistent with some recent large scale cosmological
observations (for instance the cosmic microwave background anomalies).
\end{enumerate}

Motivated by these two weak points, a wide range of
more complex generalized cosmological models has been investigated. Most of
these models reduce to \lcdm for specific values of their parameters. They
can be classified in two broad classes: Modified gravity models constitute
the one class (see for instance \cite{Capozziello:2011et}),
with the other being the scalar field dark energy that
adheres to general relativity (see
for instance \cite{Ame10,Ame10b}). Among the variety of modified
gravity theories, $f(T)$ gravity has recently gained a lot of attention.
It is based on the old formulation of the teleparallel equivalent of
general relativity (TEGR) \cite{ein28,Hayashi79,Maluf:1994ji}. In
teleparallel formulations the dynamical fields are the four linearly
independent vierbeins, while one uses the curvatureless
Weitzenb{\"{o}}ck  connection instead of the
torsionless Levi-Civita  one. Thus, one can construct the torsion tensor,
which includes all the information concerning the gravitational field, and
then by suitable contractions one can write down the corresponding Lagrangian
density $T$ \cite{Hayashi79} (assuming invariance under general
coordinate transformations, global Lorentz and parity transformations, and
requiring up to second-order terms of the torsion tensor). Finally, $f(T)$
gravity arises as a natural extension of TEGR, if one generalizes the
Lagrangian to be a function of $T$ \cite{Ferraro:2006jd,Ben09,Linder:2010py},
inspired by the well-known extension of
$f(R)$ Einstein-Hilbert action. However, the significant advantage is that
although the curvature tensor contains second-order derivatives of the
metric and thus $f(R)$ gravity gives rise to fourth-order equations which
may   lead to pathologies,  the torsion tensor includes only products of
first derivatives of the vierbeins, giving rise to second-order field equations.

Although TEGR coincides completely with general relativity both at
the background and perturbation levels, $f(T)$ gravity exhibits
novel structural and phenomenological features.
In particular, imposing a cosmological background one can extract various
cosmological solutions, consistent with the observable behavior
\cite{Ferraro:2006jd,Ben09,Linder:2010py,
Myrzakulov:2010vz,Wu:2010mn,Bengochea001,Iorio:2012cm}.
Additionally, imposing spherical
geometry one can investigate the spherical, black-hole solutions of $f(T)$
gravity \cite{Wang:2011xf}. However, we stress that although TEGR coincides
with GR,
$f(T)$ gravity does not coincide with $f(R)$ extension, but it rather
constitutes a different class of modified gravity.

One crucial question is what classes of $f(T)$ extensions are allowed
by observations. At the theoretical level, the aforementioned cosmological and spherical solutions lead to a variety of such expressions. However, taking into account
observational data, either from cosmological
\cite{Wu:2010mn,Bengochea001,Zhang:2012jsa,Cardone:2012xq} as
well as from Solar System observations \cite{Iorio:2012cm},
one can show that the deviations from TEGR must be small.

In the present work we are interested in constraining the $f(T)$ forms
using the latest cosmological data, both at the background
and perturbation levels.
In order to do so we need to define the Hubble parameter as a function of
redshift. The issue of using iterative techniques in order to treat the
Hubble expansion in $f(R)$ gravity has been proposed by Starobinsky in Ref.
\cite{Starobinsky:2007hu}. Furthermore, in a recent paper some of us
\cite{BasNes13} used a new iterative approach in order to observationally
constrain deviations of $f(R)$ models from \lcdm and general relativity. In
this context, we first showed that all known viable $f(R)$ models may be
written as perturbations around \lcdm with a deviation parameter we called
$b$ (for $b=0$ these models reduce to $\Lambda$CDM). Using a novel
perturbative iterative technique we were able to construct analytic
cosmological expansion solutions of a highly nonlinear and stiff system of
ordinary differential equations and impose cosmological observational constraints on the deviation
parameter $b$.

We also showed that the observationally viable $f(R)$ models effectively include the cosmological constant even though they were proposed as being free from a cosmological
constant in the original $f(R)$ papers \cite{Hu07,Starobinsky:2007hu}.
Inspired by our previous similar work on $f(R)$ gravity \cite{BasNes13},  we
extend it to the case of $f(T)$ gravity models and use the standard joint
likelihood analysis of the recent supernovae type Ia
data  (SnIa), the cosmic microwave background (CMB) shift parameters, the
baryonic acoustic oscillations (BAO) and the growth rate data provided by the
various galaxy surveys.  Based on these cosmological observations we identify the viable range of parameters of five previously proposed $f(T)$ models.
Additionally, comparing the resulting analytical expressions of the $f(T)$
Hubble parameter with the numerical solutions at low and intermediate
redshifts, we verify that our iterative perturbative technique is highly
accurate.

The plan of the work is as follows: In Sec. \ref{model} we briefly
discuss the main properties of the
$f(T)$ gravity, while in Sec. \ref{fTcosmology} we apply the $f(T)$
gravity in a cosmological framework, providing the relevant equations both
at the background and perturbation levels. In Sec. \ref{fTmodels} we
present and we analytically elaborate on all the $f(T)$ models of the
literature  with two parameters (out of which one is independent).
In Sec. \ref{Observationalconstraints} we impose
observational constraints, utilizing three parametrizations of the growth
index. Finally, the main conclusions are
summarized in Sec. \ref{conclusions}.

\section{$f(T)$ gravity}
\label{model}

In this section we briefly review the $f(T)$ gravitational paradigm. In this
construction the dynamical variables are the
vierbein fields ${\mathbf{e}_A(x^\mu)}$ \footnote{Throughout the manuscript,
greek indices $\mu, \nu,$... and capital Latin indices $A, B, $...
run over all coordinate and tangent space-time 0, 1, 2, 3, while lower case
latin indices (from the beginning of the alphabet) $a,b, $... and lower case
latin indices (from the middle of the alphabet) $i, j,...$, run  over
tangent-space and spatial coordinates 1, 2, 3 respectively.}.
 The vierbeins at each point $x^\mu$ of the manifold form an orthonormal
basis for the tangent space, that is $\mathbf{e}
_A\cdot%
\mathbf{e}_B=\eta_{AB}$, with $\eta_{AB}={\rm diag} (1,-1,-1,-1)$,
and they  can be expressed in terms of the
components $
e_A^\mu$ in a coordinate basis as
$\mathbf{e}_A=e^\mu_A\partial_\mu $.
Therefore, the metric tensor is obtained from the
dual vierbein through
\begin{equation}  \label{metrdef}
g_{\mu\nu}(x)=\eta_{AB}\, e^A_\mu (x)\, e^B_\nu (x).
\end{equation}
While in usual gravitational formalism  one uses the torsionless
Levi-Civita connection, in the present formulation one uses the curvatureless Weitzenb\"{o}ck connection defined as
$\overset{\mathbf{w}}{\Gamma}^\lambda_{\nu\mu}\equiv e^\lambda_A\:
\partial_\mu
e^A_\nu$ \cite{Weitzenb23}, and the corresponding torsion tensor is written as
\begin{equation}
\label{torsion2}
{T}^\lambda_{\:\mu\nu}=\overset{\mathbf{w}}{\Gamma}^\lambda_{
\nu\mu}-%
\overset{\mathbf{w}}{\Gamma}^\lambda_{\mu\nu}
=e^\lambda_A\:(\partial_\mu
e^A_\nu-\partial_\nu e^A_\mu).
\end{equation}
Furthermore, the contorsion tensor, which provides the difference
between Weitzenb\"{o}ck and Levi-Civita connections, is given by
$K^{\mu\nu}_{\:\:\:\:\rho}\equiv-\frac{1}{2}\Big(T^{\mu\nu}_{
\:\:\:\:\rho}
-T^{\nu\mu}_{\:\:\:\:\rho}-T_{\rho}^{\:\:\:\:\mu\nu}\Big)$, while for
convenience we define
$
S_\rho^{\:\:\:\mu\nu}\equiv\frac{1}{2}\Big(K^{\mu\nu}_{\:\:\:\:\rho}
+\delta^\mu_\rho
\:T^{\alpha\nu}_{\:\:\:\:\alpha}-\delta^\nu_\rho\:
T^{\alpha\mu}_{\:\:\:\:\alpha}\Big)$.
Finally, imposing coordinate, Lorentz and parity symmetries,
and the additional requirement the Lagrangian to be second
order in the
torsion tensor \cite{Hayashi79,Maluf:1994ji}, one obtains the
teleparallel Lagrangian (called ``torsion scalar'' too)
\begin{equation}
\label{torsionscalar}
T\equiv\frac{1}{4}
T^{\rho \mu \nu}
T_{\rho \mu \nu}
+\frac{1}{2}T^{\rho \mu \nu }T_{\nu \mu\rho }
-T_{\rho \mu }^{\ \ \rho }T_{\
\ \ \nu }^{\nu \mu }.
\end{equation}
Thus, in the teleparallel gravitational paradigm, all the information
concerning the gravitational field is embedded in the torsion tensor
${T}^\lambda_{\:\mu\nu} $, which produces the torsion scalar $T$
in a similar way as the curvature Riemann tensor gives rise to the Ricci
scalar in standard general relativity.

In the teleparallel equivalent of general relativity  the
action is just $T$. However, one can be inspired by the $f(R)$
extensions of general relativity  and  extend $T$ to a function $T+f(T)$.
Therefore, the corresponding action of $f(T)$ gravity
reads as
\begin{eqnarray}
\label{action00}
I = \frac{1}{16\pi G_N}\int d^4x e \left[T+f(T)\right],
\end{eqnarray}
where $e = \text{det}(e_{\mu}^A) = \sqrt{-g}$, $G_N$ is the gravitational
constant, and we use units where the light speed is equal to 1. Lastly,
TEGR and thus general
relativity is restored when $f(T)=0$, while if $f(T)=$ const we recover
general relativity with a cosmological constant.

\section{$f(T)$ cosmology}
\label{fTcosmology}

We now proceed to the cosmological application of $f(T)$ gravity. In order to construct a realistic cosmology we have to incorporate in the action
the matter and the radiation sectors respectively. Therefore, the total
action is written as
\begin{eqnarray}
\label{action11}
 I = \frac{1}{16\pi G_N }\int d^4x e
\left[T+f(T)+L_m+L_r\right],
\end{eqnarray}
where the matter and radiation Lagrangians  are assumed to correspond
to perfect fluids
with energy densities $\rho_m$, $\rho_r$ and pressures $P_m$, $P_r$
respectively.

Secondly, in order to examine a universe governed by $f(T)$ gravity, we have
to impose the usual homogeneous and isotropic geometry. Therefore,
we consider the
common choice for the vierbien form, that is,
\begin{equation}
\label{weproudlyuse}
e_{\mu}^A={\rm
diag}(1,a,a,a),
\end{equation}
which corresponds to a flat Friedmann-Robertson-Walker (FRW) background
geometry with metric
\begin{equation}
ds^2= dt^2-a^2(t)\,\delta_{ij} dx^i dx^j,
\end{equation}
with $a(t)$ the scale factor.

\subsection{Background behavior}
\label{backbehav}

Varying the action (\ref{action11}) with
respect to the vierbeins we acquire the field equations
\begin{eqnarray}\label{eom}
&&e^{-1}\partial_{\mu}(ee_A^{\rho}S_{\rho}{}^{\mu\nu})[1+f_{T}]
 +
e_A^{\rho}S_{\rho}{}^{\mu\nu}\partial_{\mu}({T})f_{TT}\ \ \ \ \  \ \ \ \  \ \
\ \ \nonumber\\
&& \ \ \ \
-[1+f_{T}]e_{A}^{\lambda}T^{\rho}{}_{\mu\lambda}S_{\rho}{}^{\nu\mu}+\frac{1}{4} e_ { A
} ^ {
\nu
}[T+f({T})] \nonumber \\
&&= 4\pi Ge_{A}^{\rho}\overset {\mathbf{em}}T_{\rho}{}^{\nu},
\end{eqnarray}
where $f_{T}=\partial f/\partial T$, $f_{TT}=\partial^{2} f/\partial T^{2}$,
and $\overset{\mathbf{em}}{T}_{\rho}{}^{\nu}$  stands for the usual
energy-momentum tensor.

Inserting the vierbein choice (\ref{weproudlyuse}) into the field equations
(\ref{eom}) we obtain the modified Friedmann equations
\begin{eqnarray}\label{background1}
&&H^2= \frac{8\pi G_N}{3}(\rho_m+\rho_r)
-\frac{f}{6}+\frac{Tf_T}{3}\\\label{background2}
&&\dot{H}=-\frac{4\pi G_N(\rho_m+P_m+\rho_r+P_r)}{1+f_{T}+2Tf_{TT}},
\end{eqnarray}
where
$H\equiv\dot{a}/a$ is the Hubble parameter, with the dot denoting
derivatives with respect to the cosmic time $t$. We mention that in order to
bring the Friedmann equations closer to their standard form, we used the relation
\begin{eqnarray}
\label{TH2}
T=-6H^2,
\end{eqnarray}
which through (\ref{torsionscalar}) arises straightforwardly for a FRW
universe.

Observing the form of the first Friedmann equation
(\ref{background1}), and comparing to the usual one, we deduce that in the
scenario at hand we obtain an effective dark energy sector of (modified)
gravitational origin. In particular, one can define the
dark energy density and pressure as \cite{Linder:2010py}
\begin{eqnarray}
&&\rho_{DE}\equiv\frac{3}{8\pi
G_N}\left[-\frac{f}{6}+\frac{Tf_T}{3}\right], \label{rhoDDE}\\
\label{pDE}
&&P_{DE}\equiv\frac{1}{16\pi G_N}\left[\frac{f-f_{T} T
+2T^2f_{TT}}{1+f_{T}+2Tf_{TT}}\right],
\end{eqnarray}
while its effective equation-of-state parameter reads:
\begin{eqnarray}
\label{wfT}
 w
=-\frac{f/T-f_{T}+2Tf_{TT}}{\left[1+f_{T}+2Tf_{TT}\right]\left[f/T-2f_{T}
\right] }.
\end{eqnarray}

In order to quantitatively elaborate the above modified Friedmann equations,
and confront them with observations, we follow the usual procedure. Firstly
we define
\begin{eqnarray}
\label{TH3}
E^{2}(z)\equiv\frac{H^2(z)} {H^2_{0}}=\frac{T(z)}{T_{0}},
\end{eqnarray}
where $T_0\equiv-6H_{0}^{2}$. Also,  we have used the redshift
$z=\frac{a_0}{a}-1$ as the independent variable and denoted by ``0'' the
current value of a quantity (in the
following we set $a_0=1$). Thus, using also that
$\rho_{m}=\rho_{m0}(1+z)^{3}$, $\rho_{r}=\rho_{r0}(1+z)^{4}$, we can
rewrite the first Friedmann equation (\ref{background1}) as
\begin{eqnarray}\label{Mod1Ez}
E^2(z,{\bf r})=\Omega_{m0}(1+z)^3+\Omega_{r0}(1+z)^4+\Omega_{F0} y(z,{\bf r})
\end{eqnarray}
with
\begin{equation}
\label{LL}
\Omega_{F0}=1-\Omega_{m0}-\Omega_{r0} \;,
\end{equation}
where $\Omega_{i0}=\frac{8\pi G \rho_{i0}}{3H_0^2}$ is the corresponding
density parameter at present. Therefore, the effect of the $f(T)$ gravity is
quantified by the function  $y(z,{\bf r})$ (normalized to
unity at   present time), which depends on $\Omega_{m0},\Omega_{r0}$, as
well as on the $f(T)$-form parameters $r_1,r_2,...$, and it is of the form
\begin{equation}
\label{distortparam}
 y(z,{\bf r})=\frac{1}{T_0\Omega_{F0}}\left[f-2Tf_T\right].
\end{equation}
According to Eq.(\ref{TH2}) the additional term (\ref{distortparam})
in the effective Friedman equation (\ref{Mod1Ez}) induced by
the $f(T)$ term is a function of the Hubble parameter only. Thus, this term is not completely arbitrary and cannot reproduce any arbitrary expansion history.
As we will show further below, the interesting point of the current analysis is that the particular range of degrees of freedom representing deviations from $\Lambda$CDM in the
context of $f(T)$ models is not favored by cosmological observations.

\subsection{Linear matter perturbations}

We now briefly discuss the linear matter perturbations of $f(T)$ gravity.
We first review the standard treatment of perturbations for general dark
energy or modified gravity scenarios. In this analysis, the extra information
is quantified by the effective Newton's gravitational constant, which appears
in the various observables such as the growth index. Thus, inserting in
these expressions the calculated effective Newton's gravitational constant
of $f(T)$ gravity, we obtain the corresponding perturbation observables of
$f(T)$ cosmology.

In the framework of any dark energy model,
including those of modified gravity (``geometrical dark energy''),
it is well known that at the subhorizon scales the dark energy component
is expected to be smooth, and thus we can
consider perturbations only on the matter component of the
cosmic fluid \cite{Dave02}.
We refer the reader to Refs.
\cite{BasNes13,Gann09,Lue04,Linder05,Stab06,Uzan07,Tsu08} for full
details of the
calculation, summarizing only the relevant results in this
section.

The basic equation which governs the behavior of the matter
perturbations in the linear regime is written as
\begin{equation}
\label{odedelta}
\ddot{\delta}_{m}+ 2H\dot{\delta}_{m}=4 \pi G_{\rm eff} \rho_{m} \delta_{m},
\end{equation}
where $\rho_{m}$ is the matter density and $G_{\rm eff}(a)=G_{N} Q(a)$, with $G_{N}$ denoting Newton's gravitational constant. That is, the effect of the modified  gravity at the
linear perturbation level is reflected in an effective Newton's
gravitational constant $G_{\rm eff}(a)$, which in general is evolving.
Finally, in the above analysis it has been found that
$\delta_{m}(t) \propto D(t)$, where $D(t)$ is the linear growth
factor normalized to unity at present time.

In the case of general-relativity-based scalar-field dark energy models, we
obviously have $G_{\rm eff}(a)=G_{N}$ [that is $Q(a)=1$] and
therefore (\ref{odedelta}) reduces to the usual time-evolution
equation for the mass density contrast \cite{Peeb93}.
Moreover, in the case of the usual $\Lambda$ cosmology, one
can solve  (\ref{odedelta}) analytically in order to obtain
the growth factor   \cite{Peeb93}
\begin{equation}
\label{eq24}
D_{\Lambda}(z)=\frac{5\Omega_{m0}
  E_{\Lambda}(z)}{2}\int^{+\infty}_{z}
\frac{(1+u)du}{E^{3}_{\Lambda}(u)},
\end{equation}
where
\begin{equation}
E_{\Lambda}(z)=\left[ \Omega_{m0}(1+z)^{3}+1-\Omega_{m0}\right]^{1/2}
\end{equation}
in the matter dominated era.

In general for either dark energy
or modified gravity scenarios, a useful tool that
simplifies the numerical calculations significantly is the
growth rate of clustering
\cite{Peeb93}
\begin{equation}
\label{fzz221}
F(a)=\frac{d\ln \delta_{m}}{d\ln a}\simeq \Omega^{\gamma}_{m}(a),
\end{equation}
where $\gamma$ is the growth index, which is general evolving. The growth
index is very important since it can be used to distinguish between general
relativity and modified gravity on cosmological scales. Indeed,  for a
constant dark energy equation of state parameter $w$,
dark energy scenarios in the framework of general relativity the growth index is
well approximated by $\gamma \simeq \frac{3(w-1)}{6w-5}$
\cite{Silv94,Wang98,Lue04,Linder2007,Nes08},
which reduces to $\approx 6/11$ for the concordance $\Lambda$ cosmology ($w=-1$).
On the other hand,for the braneworld model of Dvali,  Gabadadze and Porrati
\cite{Dvali2000} the growth index becomes $\gamma \approx 11/16$
\cite{Linder2007,Gong10,Wei08,Fu09}, for some $f(R)$ gravity models one
acquires $\gamma \simeq 0.415-0.21z$ for various parameter values
\cite{Gann09,Tsu09}, while for Finsler-Randers cosmology we have $\gamma
\approx 9/14$  \cite{Bastav13}.

Generally, combining Eq. (\ref{odedelta}) with the first equality of
 (\ref{fzz221}) we obtain
\begin{equation}
\label{fzz222}
a\frac{dF(a)}{da}+F(a)^{2}+X(a)F(a)
= \frac{3}{2}\Omega_{m}(a)Q(a) \;,
\end{equation}
with
\begin{equation}
\label{xxa}
X(a)=\frac{1}{2}-\frac{3}{2}w(a)
\left[ 1-\Omega_{m}(a)\right] ,
\end{equation}
where we have used that   \cite{Ame10,Ame10b,BasNes13,Saini00}
\begin{equation}
\label{eos222}
w(a)=\frac{-1-\frac{2}{3}a\frac{{d\rm lnE}}{da}}
{1-\Omega_{m}(a)} \;,
\end{equation}
\begin{equation}
\label{ddomm}
\Omega_{m}(a)=\frac{\Omega_{m0}a^{-3}}{E^{2}(a)} \,,
\end{equation}
and
\begin{equation}
\label{domm}
\frac{d\Omega_{m}(a)}{da}=
\frac{3}{a}w(a)\Omega_{m}(a)\left[1-\Omega_{m}(a)\right]\;.
\end{equation}

Concerning the functional form of the growth index we consider various
situations. The simplest one is to use a constant
growth index (hereafter $\Gamma_{0}$ model). If we allow $\gamma$ to be
a function of redshift then Eq.(\ref{fzz222}) can be expressed in terms of $\gamma=\gamma(z)$
and it is given by
{\small{
\begin{equation}
\label{Poll}
-(1+z)\gamma^{\prime}{\rm ln}(\Omega_{m})+\Omega_{m}^{\gamma}+
3w(1-\Omega_{m})\left(\gamma-\frac{1}{2}\right)+\frac{1}{2}=\frac{3}{2}
Q\Omega_ { m } ^ { 1-\gamma},
\end{equation}}}
where prime denotes derivative with respect to redshift.
Writing the above equation at the present epoch ($z=0$) we have
\begin{eqnarray}
\label{Poll1}
&&-\gamma^{\prime}(0){\rm
ln}(\Omega_{m0})+\Omega_{m0}^{\gamma(0)}\nonumber\\
&&    +
3w_{0}(1-\Omega_{m0})\left[\gamma(0)-\frac{1}{2}\right]+\frac{1}{2}=\frac { 3
} { 2 } Q_ { 0 } \Omega_{m0}^{1-\gamma(0)}, \ \ \
\end{eqnarray}
where $Q_{0}=Q(z=0)$ and $w_{0}=w(z=0)$.

In this work we consider some well known $\gamma(z)$ functional forms
(see \cite{Pol,Bel12,DP11,Ishak09}). These parametrizations are
\begin{equation}
\gamma(z)=\left\{ \begin{array}{cc}
       \gamma_{0}, &
       \mbox{$\Gamma_{0}$ model}\\
       \gamma_{0}+\gamma_{1}z, &
       \mbox{$\Gamma_{1}$ model}\\
       \gamma_{0}+\gamma_{1}(1-a),&\ \mbox{$\Gamma_{2}$ model.}
       \end{array}
        \right.
\end{equation}
 Inserting the $\Gamma_{1-2}$ formulas into Eq.(\ref{Poll1}) one can easily write
the parameter $\gamma_{1}$ in terms of $\gamma_{0}$:
\begin{equation}
\label{Poll2}
\gamma_{1}=\frac{\Omega_{m0}^{\gamma_{0}}+3w_{0}(\gamma_{0}-\frac{1}{2})
(1-\Omega_{m0})-\frac{3}{2}Q_{0}\Omega_{m0}^{1-\gamma_{0}}+\frac{1}{2}  }
{\ln  \Omega_{m0}}\;.
\end{equation}
Finally, we would like to stress that the $\Gamma_{1}$
parametrization is valid only at relatively low redshifts $0\le z \le 0.5$.
Therefore, in the statistical analysis presented below we utilize a constant
growth index, namely, $\gamma=\gamma_{0}+0.5\gamma_{1}$ for $z>0.5$.

Since we now have the general perturbation formulation, we just need to
insert $G_{\rm eff}(a)$, or equivalently $Q(a)$, of $f(T)$ gravity in the
above relations. Unlike the $f(R)$ gravity, the effective Newton's parameter
in $f(T)$ gravity is not affected by the scale but rather it
takes the following form \cite{Zheng:2010am}:
\begin{eqnarray}
\label{Geff}
Q(a)=\frac{G_{\rm eff}(a)}{G_{N}}=\frac{1}{1+f_{T}},
\end{eqnarray}
as it arises from the complete perturbation analysis \cite{Chen001}.
The above can be understood, as it was shown in Ref.\cite{Myrzakulov},
from the fact that the $f(T)$ cosmological scenario
can be rewritten the as K-essence model which implies that since
we remain at the Jordan frame we do not expect to have a $k$ dependence
in the effective Newton's parameter and thus in
the growth factor. However, doing a similar exercise for the $f(R)$ gravity [see Eqs.
(8) (10) in Ref.\cite{Ferraro}] one can easily find that it corresponds to a scalar-tensor theory, i.e. a nonminimally coupled scalar field which obviously induces a $k$ dependence in the matter density perturbations. Therefore, in the rest of the work we apply the above analysis in the case of $f(T)$, that is, with $Q(a)$ given by (\ref{Geff}).

\section{Specific $f(T)$ models and the deviation from $\Lambda$CDM}
\label{fTmodels}

In this section we review all the specific $f(T)$ models that have appeared
in the literature,
with two parameters out of which one is independent. We  calculate the
function $y(z,{\bf r})$ using (\ref{distortparam}) and their
$G_{\rm eff}(a)$ using  (\ref{Geff}).   We quantify the deviation of the
function $y(z,{\bf r})$ from its \lcdm value (constant) through a distortion
parameter $b$. The considered models are as follows.

\begin{enumerate}
\item The power-law model of Bengochea and Ferraro
(hereafter $f_{1}$CDM) \cite{Ben09}, with
\begin{equation}
f(T)=\alpha (-T)^{b},
\end{equation}
where $\alpha$ and $b$ are the two model parameters.
Substituting this $f(T)$ form
into the modified Friedmann equation (\ref{background1}) at present, we
obtain
\begin{eqnarray}
\alpha=(6H_0^2)^{1-b}\frac{\Omega_{F0}}{2b-1},
\end{eqnarray}
while (\ref{distortparam})
gives
\begin{equation}
\label{yLL}
y(z,b)=E^{2b}(z,b) \;.
\end{equation}
Additionally, the effective Newton's constant from  (\ref{Geff})
becomes
\begin{equation}
G_{\rm eff}(z)=\frac{G_{N}}{1+ \frac{b\Omega_{F0}} {(1-2b)E^{2(1-b)}}}\;.
\end{equation}

It is evident that for $b$ strictly equal to zero the $f_{1}$CDM model
reduces to $\Lambda$CDM cosmology, namely
$T+f(T)=T-2\Lambda$ (where $\Lambda=3\Omega_{F0}H_{0}^{2}$,
$\Omega_{F0}=\Omega_{\Lambda 0}$),
while for  $b=1/2$ it reduces to the Dvali, Gabadadze and Porrati (DGP)
ones \cite{Dvali2000}. Note that in order to
obtain an accelerating expansion, it is required that $b<1$.

\item The Linder model (hereafter $f_{2}$CDM) \cite{Linder:2010py}
\begin{eqnarray}
f(T)=\alpha T_{0}(1-e^{-p\sqrt{T/T_{0}}}),
\end{eqnarray}
with $\alpha$ and $p$  the two model parameters. In this case from
(\ref{background1})  we find that
\begin{eqnarray}
\alpha=\frac{\Omega_{F0}}{1-(1+p)e^{-p}}\;,
\end{eqnarray}
and from (\ref{distortparam}) we acquire
\begin{equation}
\label{yLL1}
y(z,p)=\frac{1-(1+pE)e^{-pE}}{1-(1+p)e^{-p}} \;,
\end{equation}
while from  (\ref{Geff}) we obtain
\begin{equation}
G_{\rm eff}(z)=\frac{G_{N}}{1+ \frac{\Omega_{F0}p~e^{-pE}}
{2E[1-(1+p)e^{-p}]}   } \;.
\end{equation}

Thus, for $p \rightarrow +\infty$ the $f_{2}$CDM reduces
to  $\Lambda$CDM cosmology, since
\begin{equation}
\lim_{p\rightarrow +\infty}[T+f(T)]=T-2\Lambda \;.
\end{equation}
The parameter $p$ of the present $f_{2}$CDM model
has a different interpretation comparing to $b$ for the $f_{1}$CDM
model, since the two models are obviously different.
However, since in the limiting case they both reduce to
$\Lambda$CDM paradigm, we can rewrite the present  $f_{2}$CDM model
replacing  $p=1/b$. In this case  (\ref{yLL1}) leads to
\begin{equation}
\label{yLL2}
y(z,b)=\frac{1-(1+\frac{E}{b})e^{-E/b}}{1-(1+\frac{1}{b})e^{-1/b}},
\end{equation}
which indeed tends to unity for
$b \rightarrow 0^{+}$.

\item Motivated by  exponential $f(R)$ gravity \cite{Linn1}, one can
construct the following   $f(T)$ model (hereafter $f_{3}$CDM):
\begin{eqnarray}
f(T)=\alpha T_{0}(1-e^{-pT/T_{0}}),
\end{eqnarray}
with $\alpha$ and $p$  the two model parameters.
In  this case we obtain
\begin{eqnarray}
\alpha=\frac{\Omega_{F0}}{1-(1+2p)e^{-p}} \;,
\end{eqnarray}
\begin{equation}
\label{Mod223}
y(z,p)=\frac{1-(1+2pE^{2})e^{-pE^{2}}}{1-(1+2p)e^{-p}} \;.
\end{equation}
and
\begin{equation}
G_{\rm eff}(z)=\frac{G_{N}}{1+ \frac{\Omega_{F0}p~ e^{-pE^{2}}}
{1-(1+2p)e^{-p}}   } \;.
\end{equation}
Similarly to the previous case we can rewrite  $f_{3}$CDM model using
 $p=1/b$, obtaining
\begin{equation}
\label{Mod224}
y(z,b)=\frac{1-(1+\frac{2E^{2}}{b})e^{-E^{2}/b}}{1-(1+\frac{2}{b})e^{-1/b}} \;.
\end{equation}
Again, we see that for $p \rightarrow +\infty$, or equivalently for $b
\rightarrow 0^{+}$, the $f_{3}$CDM model tends to the $\Lambda$CDM cosmology.

\item The Bamba \textit{et al.} logarithmic model
(hereafter $f_{4}$CDM) \cite{Bamba}
\begin{eqnarray}
f(T)=\alpha T_{0} \sqrt{\frac{T}{qT_{0}}}\;
{\rm ln}\left( \frac{qT_{0}}{T}\right )
\end{eqnarray}
with $\alpha$ and $q$  the two model parameters.
In this case we obtain
\begin{eqnarray}
\alpha=\frac{\Omega_{F0} \sqrt{q}}{2} \;,
\end{eqnarray}
\begin{equation}
\label{Mod44}
y(z)=E(z) \;,
\end{equation}
and
\begin{eqnarray}
\label{GM}
G_{\rm
eff}(z)=\frac{G_{N}}{1+\frac{\Omega_{F0}}{2E}\left[\ln\left(\frac{\sqrt{q}}{E
}\right) -1\right ] }.
\end{eqnarray}
The fact that the distortion function does not depend on the model
parameters, allows us to write (\ref{Mod1Ez}) as
\begin{eqnarray}
E(z)&=&\frac{1}{2}
\sqrt{\Omega^{2}_{F0}+4\left[\Omega_{m0}(1+z)^{3}+
\Omega_{r0}(1+z)^{4}\right]}\nonumber\\
&+&\frac{\Omega_{F0}}{2}.
\end{eqnarray}
Interestingly enough, from the above relation we deduce that at the
background level the $f_{4}$CDM model coincides with the flat DGP one (with
$\Omega_{F0}=\Omega_{DGP}$), which implies that the two nonstandard gravity
models are cosmologically equivalent as far as the cosmic expansion is
concerned,  in spite of the fact that the two models have
a completely different geometrical basis. At
the perturbative level, however, we do expect to find
differences between $f_{4}$CDM and DGP, since $G_{\rm eff}(z)$ evolves
differently in two models [in flat DGP gravity we have
$\frac{G_{\rm
eff}(z)}{G_{N}}=\frac{2+4\Omega^{2}_{m}(z)}{3+3\Omega^{2}_{m}(z)}$].

Notice that this model does not give $\Lambda$CDM
cosmology for any value of its parameters. However, in this work we are
interested in the viable $f(T)$,  in the sense that these $f(T)$ models
can describe the matter and dark energy eras as well as they are
consistent with the observational data (including Solar System tests), and finally they have stable perturbations. Although these necessary analysis have not yet been performed for all the above $f(T)$ models, a failure of a particular model to pass one of these
is enough to exclude it. Therefore, since the present $f_{4}$CDM model
coincides with DGP at the background level, it inherits its disadvantages
concerning the confrontation with observations. Thus, as anticipated from
previous studies \cite{Fang:2008kc}, we verify in the following section that
this model is nonviable when tested using the latest cosmological observations.

\item The hyperbolic-tangent model (hereafter $f_{5}$CDM) \cite{Wu:2011}
\begin{eqnarray}
f(T)=\alpha(-T)^{n}{\rm tanh}\left( \frac{T_{0}}{T}\right)
\end{eqnarray}
with $\alpha$ and $n$  the two model parameters.
In this case we obtain
\begin{eqnarray}
\alpha=-\frac{\Omega_{F0}(6H_{0})^{1-n}}{\left[ 2{\rm sech}^{2}(1)+(1-2n){\rm
tanh}(1)\right]} \;,
\end{eqnarray}
{\small{
\begin{eqnarray}
y(z,n)=E^{2(n-1)}\frac{2{\rm
sech}^{2}\left(\frac{1}{E^2}\right)+(1-2n)E^2{\rm
tanh}\left(\frac{1}{E^2}\right)}{2{\rm
sech}^{2}(1)+(1-2n){\rm tanh}(1)}
\end{eqnarray}}}
and
\begin{eqnarray}
\label{GM}
G_{\rm
eff}(z)=\frac{G_{N}}{1+\frac{\Omega_{F0} E^{2(n-2)}   \left[n E^2   {\rm
tanh}\left(\frac{1}{E^2}\right)   -  {\rm
sech}^2\left(\frac{1}{E^2}\right)
\right]}{2{\rm sech}^{2}(1)+(1-2n){\rm
tanh}(1)}}.
\end{eqnarray}

The $f_{5}$CDM model does not give $\Lambda$CDM cosmology for any value of
its parameters. However, as we show in the next section, this model is in mild
tension with the data as it has a best fit $\chi_{min}^2=(579.583, 580.723,
578.027)$ for the $\Gamma_0$, $\Gamma_1$ and $\Gamma_2$ growth rate
parameterizations respectively, which is significantly larger than that of $f_{1-3}$CDM
and $\Lambda$CDM models respectively (see Table \ref{tab:growth1}).
Additionally the current $f(T)$ model has one more free parameter. For the
reasons developed above we consider it as nonviable
(see also akaike information criterion (AIC) test in Table I.)
\end{enumerate}

The above five $f(T)$ forms are the ones that have been used in the
literature of $f(T)$ cosmology, possessing up to two parameters,
out of which one is independent.
Clearly, in principle one could additionally consider their combinations
too; however, the appearance of many free parameters would be a significant
disadvantage. Therefore, in the present work we focus only on these five
elementary \textit{Ans\"{a}tze}.

As we showed, for the first three the  distortion parameter measures the
smooth deviation from the $\Lambda$CDM model. The other two models do not
have $\Lambda$CDM cosmology as a limiting case; however, as we show in the next section, they
are in tension
with observations. Thus, in the
rest of this section we focus on the first three models, namely on
$f_{1-3}$CDM ones.

\begin{figure*}[ht]
\centering
\vspace{0cm}\rotatebox{0}{\vspace{0cm}\hspace{0cm}\resizebox{0.49\textwidth}{!}{\includegraphics{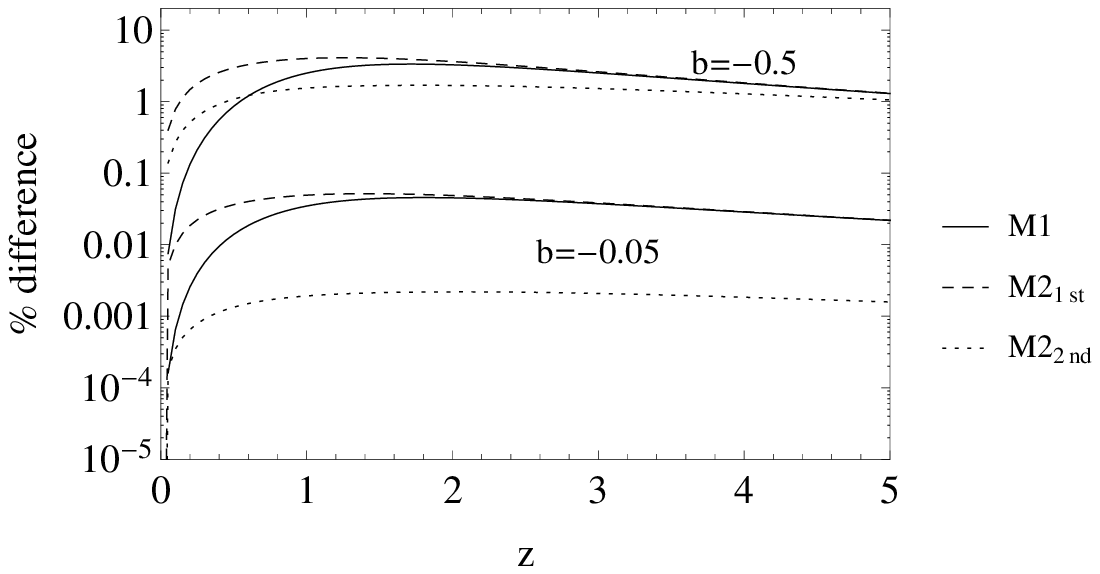}}}
\vspace{0cm}\rotatebox{0}{\vspace{0cm}\hspace{0cm}\resizebox{0.48\textwidth}{!}{\includegraphics{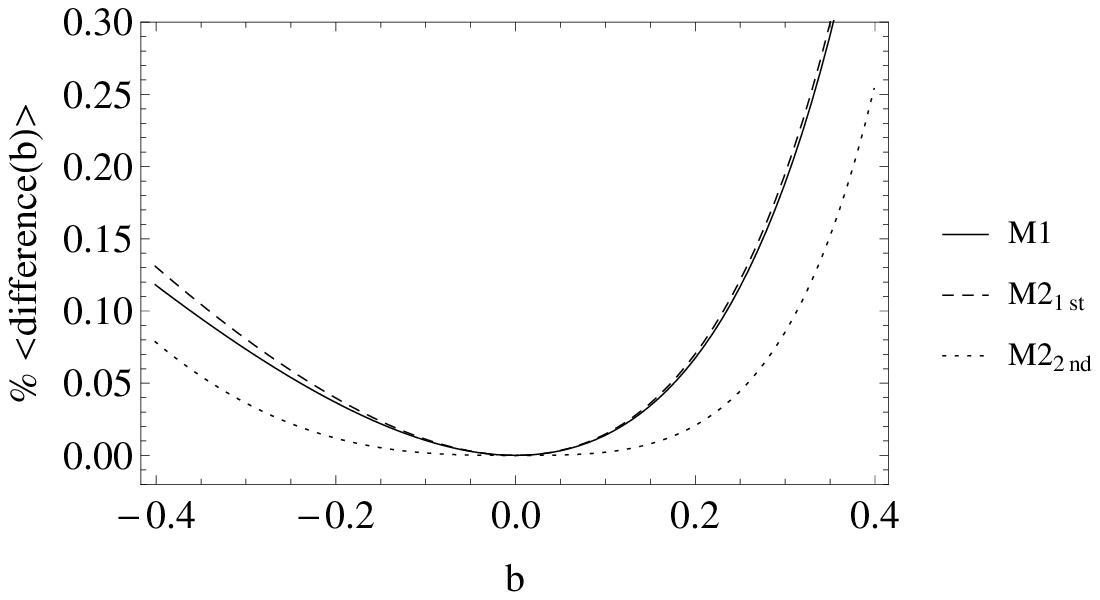}}}
\caption{Left: The percent difference$(z,b)$ between the numerical solution of Eqs.~(\ref{Mod1Ez}) and (\ref{yLL}) and the analytical approximations of Eqs.~(\ref{approxM1}) and (\ref{approxM2}) as a function of $z$, for various values of the parameter $b$ for both methods $M_1$ (at first and second order) and $M_2$. Right: The average percent difference $\langle\textrm{difference}(b)\rangle$ between the numerical solution of Eqs.~(\ref{Mod1Ez}) and (\ref{yLL}) and the analytical approximations of Eqs.~(\ref{approxM1}) and (\ref{approxM2}) as a function of the parameter $b$. In this case, the average over the redshift is taken in the range $z\in[0,100]$. \label{diffs}}
\end{figure*}

Having performed the above elaboration of various $f(T)$ models, we can now
follow the procedure and iterative techniques of Basilakos, Nesseris and Perivolaropoulos
\cite{BasNes13}, in which we have shown that all the observationally viable
$f(R)$ parameterizations can be expressed as perturbations deviating from
$\Lambda$CDM cosmology.

For the $f_1$CDM model there are two different, but complementary,  ways we
can find analytical approximations for the Hubble parameter. The first method
involves doing a Taylor expansion of $E^2(z,b)$ around $b=0$, while in the
second we perform the Taylor expansion in the modified Friedman equation
directly. Below, we briefly review and test both methods, called $M_1$ and
$M_2$ respectively.

First, from (\ref{Mod1Ez}) with (\ref{yLL}) we can write explicitly  the
Hubble parameter for the  $f_1$CDM model as
\be
E^2(z,b)=\Omega_{m0} (1+z)^3+\Omega_{r0} (1+z)^4+\Omega_{F0}\;y(z,b)\label{modfriedf1},
\ee
where
\be
y(z,b)=E^{2b}(z,b)\label{yzb}.
\ee
Obviously, in Eq.~(\ref{modfriedf1}) if we set $b$ strictly equal to zero then we get the Hubble parameter for the $\Lambda$CDM model
\be
E^2(z,0)=\Omega_{m0} (1+z)^3+\Omega_{r0} (1+z)^4+\Omega_{F0}\equiv E^2_\Lambda(z)\label{friedlcdm}.
\ee
Now, performing a Taylor expansion, up to second order, on $E^2(z,b)$ around
$b=0$ and with the help of (\ref{modfriedf1}) we arrive at
\begin{widetext}
\bea
E^2(z,b)&=&E^2(z,0)+\frac{dE^2(z,b)}{db}|_{b=0} b+ \frac{d^2E^2(z,b)}{db^2}|_{b=0} \frac{b^2}{2} +...=\nn \\
&=&E^2_\Lambda(z)+\Omega_{F0}\frac{dy(z,b)}{db}|_{b=0}
b+\Omega_{F0}\frac{d^2y(z,b)}{db^2}|_{b=0} \frac{b^2}{2}+\cdots.
\eea
\end{widetext}

The terms involving the derivatives of $y(z,b)$ can readily be calculated
from Eq.~(\ref{yzb}) as
\be
\frac{dy(z,b)}{db}=2E(z,b)^{2 b} \left\{\frac{b }{E(z,b)} \frac{dE(z,b)}{db}+ \ln
\left[E(z,b)\right]\right\},\\
\ee
and evaluating the above equation for $b=0$ we have
\be
\frac{dy(z,b)}{db}|_{b=0}=2\ln\left[E(z,0)\right]=\ln
\left[E^2_\Lambda(z)\right].\\
\ee
Similarly for the second derivative term we have
\be
\frac{d^2y(z,b)}{db^2}|_{b=0}=\frac{2\Omega_{F0}\ln\left[E^2_\Lambda(z)
\right]} { E^2_\Lambda(z)}
+\ln\left[E^2_\Lambda(z)\right]^2.
\ee
Thus, the Taylor expansion up to second order for the first method $M_1$
becomes
\bea
&&E^2(z,b)=E^2_\Lambda(z)+\Omega_{F0}\ln\left[E^2_\Lambda(z)\right]
\;b\nn\\
&&\ \ \ \   +
\Omega_{F0}\left\{\frac{2\Omega_{F0}\ln\left[E^2_\Lambda(z)\right]}{
E^2_\Lambda(z) }
+\ln\left[E^2_\Lambda(z)\right]^2\right\}\frac{b^2}{2}+\cdots.
\nn\\
\label{approxM1}
\eea

The second method $M_2$ involves performing a Taylor expansion in the modified Friedman
equation (\ref{modfriedf1}) directly. For the details in this case we refer the interested
reader to the Appendix and just present the result here:
\be
E^2(z,b)=-b~\Omega_F~\mathcal{W}_k\left(-\frac{e^{-\frac{E_{\Lambda
}(z){}^2}{b~\Omega _F}}}{b\;\Omega _F}\right),
\label{approxM2}
\ee
where $\mathcal{W}_k(\omega)$ is the Lambert function defined via
$\omega\equiv\mathcal{W}_k(\omega)e^{\mathcal{W}_k(\omega)}$ for all
complex numbers $\omega$. The Lambert function has branch-cut discontinuities, so the
different branches are indicated by the integer $k$. Our solution has $k=0$ (the principal
branch) for $b\leq0$ and $k=-1$ for $b>0${}.\footnote{The Lambert function $\mathcal{W}_k(\omega)$ is defined in \textsc{Mathematica} as ProductLog$\[k,\omega\]$ and can be evaluated to arbitrary precision for integer values of $k$ and real or complex values of $\omega$.}.

In order to examine the accuracy of the approximations of  (\ref{approxM1})
and (\ref{approxM2}), we calculate the average percent deviation from the
exact numerical solution of (\ref{modfriedf1}), defined as
\begin{equation}
\left<\textrm{\
difference}(b)\right>~=~\left<100\cdot\left(1-\frac{E^2_{approx}(z,b) }{E^2_{numeric}(z,b)}\right)\right>,
\end{equation}
where the average is taken over redshifts in the range $z\in[0,100]$. In
Fig.~\ref{diffs} we show the corresponding results. In particular, on the
left plot we show the percent difference between the numerical solution  of
Eqs.~(\ref{Mod1Ez}) and (\ref{yLL}) and the analytical approximations of
Eqs.~(\ref{approxM1}) and (\ref{approxM2}) as a function of $z$, for various
values of the parameter $b$ for both methods $M_1$, at first (dashed line)
and second order (dotted line) and $M_2$ (solid black line). As it can be
seen, at redshifts $z\lesssim2$ method $M_2$ is significantly better than the
first-order $M_1$, but overall, obviously the second-order method $M_2$
is much better than the other two.

On the right plot we show the average percent difference
$\langle\textrm{difference}(b)\rangle$ between the numerical solution of
Eqs.~(\ref{Mod1Ez}) and (\ref{yLL}) and the analytical approximations of
Eqs.~(\ref{approxM1}) and  (\ref{approxM2}) as a function of the parameter
$b$. In this case, the average over the redshift is taken in the range
$z\in[0,100]$. Clearly, on average the second-order method is significantly
better than the other two methods, the first-order $M_1$ and the $M_2$.
Thus, we conclude that the second order series expansion of
Eq.~(\ref{approxM1}) around $\Lambda$CDM for the $f_{1}$CDM model is a very
good approximation, especially for realistic values of the parameter $b$.

Unfortunately, for the $f_2$CDM and  $f_3$CDM models it is not possible to
analytically obtain similar expressions, due to the presence of terms like
$\sim e^{-1/b}$, which do not admit a Taylor expansion around $b\sim0$.
However, as mentioned earlier, they both have the $\Lambda$CDM model as a
limit for $b\rightarrow0^+$.

\section{Observational constraints}
\label{Observationalconstraints}

In this section we perform a complete and detailed observational analysis of
the above five $f(T)$ models. In particular, we implement a joint
statistical analysis with the appropriate Akaike information criterion
\cite{Akaike1974}, involving the latest expansion data (SnIa
\cite{Suzuki:2011hu}, BAO \cite{Blake:2011en,Perc10} and the 9-year WMAP CMB
shift parameter \cite{Hinshaw:2012fq}) and the growth data (as collected by
\cite{BasNes13}). The likelihood analysis, the Akaike information criterion,
the expansion data, the growth data and the corresponding covariances can be
found in Table I and Sec. IV of our previous work \cite{BasNes13}.
Moreover, we mention that since in order to deal with the growth data we need
to know the value of $\sigma_{8}$, which is the rms mass fluctuation on
$R_{8}=8 h^{-1}$ Mpc scales at redshift $z=0$, we treat $\sigma_{8}$
either as $\sigma_{8}=0.8$ or as a free parameter. This analysis is
significantly improved, comparing to previous observational constraining of
$f(T)$ gravity  \cite{Zhang:2012jsa,Bamba,Wu:2010mn,Wu:2011}.

Let us now provide a presentation of our statistical results.
In Table I we give the resulting best fit parameters for the various $f(T)$
models under study (we impose here $\sigma_{8}=0.8$), in which we also show the
corresponding quantities for $\Lambda$CDM for comparison.

It is clear that utilizing the combination of the most recent
growth data set with the expansion cosmological data, we can put tight
constraints on $(\Omega_{m},\gamma)$. In all cases the best fit value
$\Omega_{m}=0.272 \pm 0.003$ is in a very good agreement with the one found
by WMAP9+SPT+ACT, that is, $\Omega_{m}=0.272$ \cite{Hinshaw:2012fq}.
\begin{table*}[!]
\tabcolsep 4.5pt
\vspace{1mm}
\begin{tabular}{ccccccccc} \hline \hline
Exp. model & Param. model & $\Omega_{m0}$ & $b$ & $\gamma_{0}$& $\gamma_{1}$&
$\chi_{min}^{2}$ &${\rm AIC}$& $|\Delta$AIC$|$ \vspace{0.05cm}\\ \hline
             &$\Gamma_{0}$& $0.272\pm0.003$ &  & $0.597\pm 0.046$& $0$             &574.227& 578.227 & 0     \vspace{0.01cm}\\
$\Lambda$CDM &$\Gamma_{1}$& $0.272\pm0.003$ &  & $0.567\pm 0.066$& $0.116\pm0.191$ &573.861& 579.861 & 1.634 \vspace{0.01cm}\\
             &$\Gamma_{2}$& $0.272\pm0.003$ &  & $0.561\pm 0.068$& $0.183\pm0.269$ &573.767& 579.767 & 1.540 \vspace{0.45cm}\\
                         &$\Gamma_{0}$& $0.274\pm0.008$ & $-0.017\pm0.083$&$0.602\pm 0.052$& 0                &574.203& 580.203& 1.976\vspace{0.01cm}\\
$f_{1}$CDM \cite{Ben09}: &$\Gamma_{1}$&$0.275\pm0.008$ & $-0.029\pm0.088$ &$0.558\pm 0.067$& $0.187\pm 0.205$ &573.817& 581.817& 3.590\vspace{0.01cm}\\
                         &$\Gamma_{2}$& $0.275\pm0.008$ & $-0.030\pm0.089$&$0.564\pm 0.069$& $0.213\pm 0.287$ &573.640& 581.640& 3.413\vspace{0.45cm}\\
                                 &$\Gamma_{0}$&$0.272\pm0.004$ & $0.121\pm 0.184 $&$0.596\pm0.047$& 0              &$574.250$& 580.250& 2.023\vspace{0.01cm}\\
$f_{2}$CDM \cite{Linder:2010py}: &$\Gamma_{1}$&$0.272\pm0.003$ & $0.086\pm 0.301 $&$0.566\pm0.066$& $0.116\pm0.191$&$573.863$& 581.863& 3.636\vspace{0.01cm}\\
                                 &$\Gamma_{2}$&$0.272\pm0.003$ & $0.078\pm 0.375 $&$0.561\pm0.068$& $0.183\pm0.269$&$573.768$& 581.768& 3.541\vspace{0.45cm}\\
                         &$\Gamma_{0}$& $0.273\pm0.003$ & $0.097\pm0.155$&$0.597\pm0.046$& 0              &$574.223$& 580.223& 1.996\vspace{0.01cm}\\
$f_{3}$CDM \cite{Linn1}: &$\Gamma_{1}$& $0.273\pm0.003$ & $0.010\pm0.324$&$0.570\pm0.067$&$0.099\pm0.192$ &$573.852$& 581.852& 3.625\vspace{0.01cm}\\
                         &$\Gamma_{2}$& $0.273\pm0.003$ & $0.024\pm0.183$&$0.562\pm0.068$& $0.185\pm0.269$&$573.749$& 581.749& 3.522\vspace{0.45cm}\\
                         &$\Gamma_{0}$& $0.202\pm0.002$ &  & $0.417\pm0.031$&0                 &$704.481$& 708.481 & 130.254\vspace{0.01cm}\\
$f_{4}$CDM \cite{Bamba}: &$\Gamma_{1}$& $0.202\pm0.002$ &  & $0.468\pm0.053$& $-0.171\pm0.136$ &$702.865$& 708.865 & 130.638\vspace{0.01cm}\\
                         &$\Gamma_{2}$& $0.202\pm0.002$ &  & $0.467\pm0.052$&$-0.224\pm0.134$  &$703.047$& 709.047 & 130.820\vspace{0.01cm}\\
\\
                           &$\Gamma_{0}$& $0.283\pm0.006$ & $0.226\pm0.066$ & $0.567\pm0.049$&0                &$579.583$& 585.583 & 7.356\vspace{0.01cm}\\
$f_{5}$CDM \cite{Wu:2011}: &$\Gamma_{1}$& $0.277\pm0.006$ & $0.298\pm0.049$ & $0.550\pm0.065$& $0.099\pm0.191$ &$580.723$& 588.723 & 10.496\vspace{0.01cm}\\
                           &$\Gamma_{2}$& $0.287\pm0.007$ & $0.193\pm0.074$ & $0.570\pm0.070$&$0.263\pm0.298$  &$578.027$& 586.027 & 7.800\vspace{0.01cm}\\
\hline\hline
\end{tabular}
\caption[]{Statistical results of the overall likelihood analysis: The
first column indicates the $f(T)$ model, the second column
the $\gamma(z)$ parametrizations appearing in Sec. \ref{backbehav}, the
third and fourth columns provide the $\Omega_{m0}$ and $b$ best
values, and the fifth and sixth columns show the $\gamma_{0}$ and
$\gamma_1$ best fit values. In all cases we have used $\sigma_{8}=0.8$. The
last three columns present the goodness-of-fit statistics
($\chi^{2}_{min}$, AIC and $|\Delta$AIC$|=|{\rm AIC}_{\Lambda}-{\rm
AIC}_{f(T)}|$). All the error estimates come from the inverse of the Fisher
matrix, called the covariance matrix, and are by definition symmetric.\label{tab:growth1}}
\end{table*}

In particular, we find the following

\begin{figure*}[ht]
\centering
\vspace{0cm}\rotatebox{0}{\vspace{0cm}\hspace{0cm}\resizebox{0.32\textwidth}{!}{\includegraphics{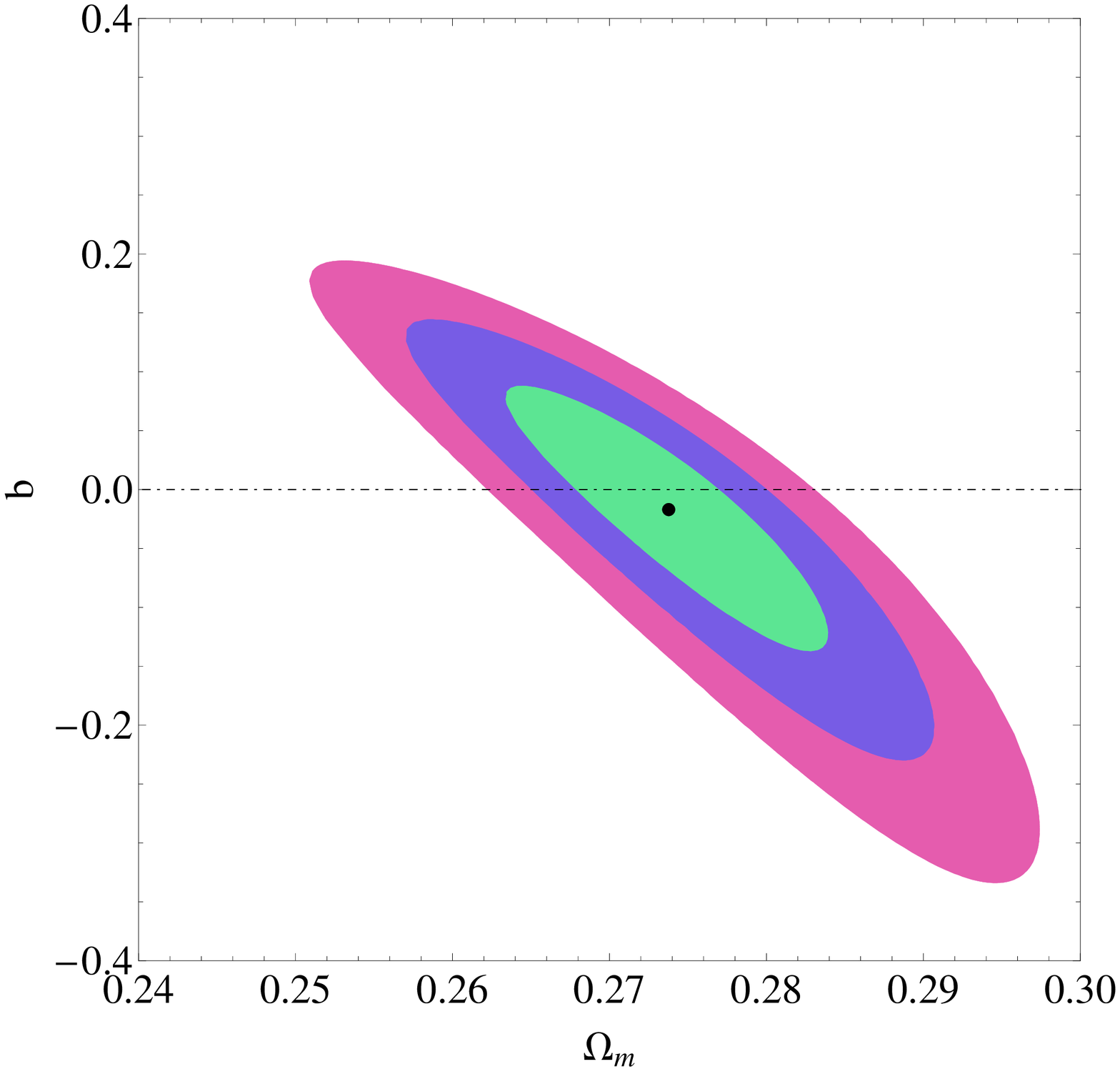}}}
\vspace{0cm}\rotatebox{0}{\vspace{0cm}\hspace{0cm}\resizebox{0.31\textwidth}{!}{\includegraphics{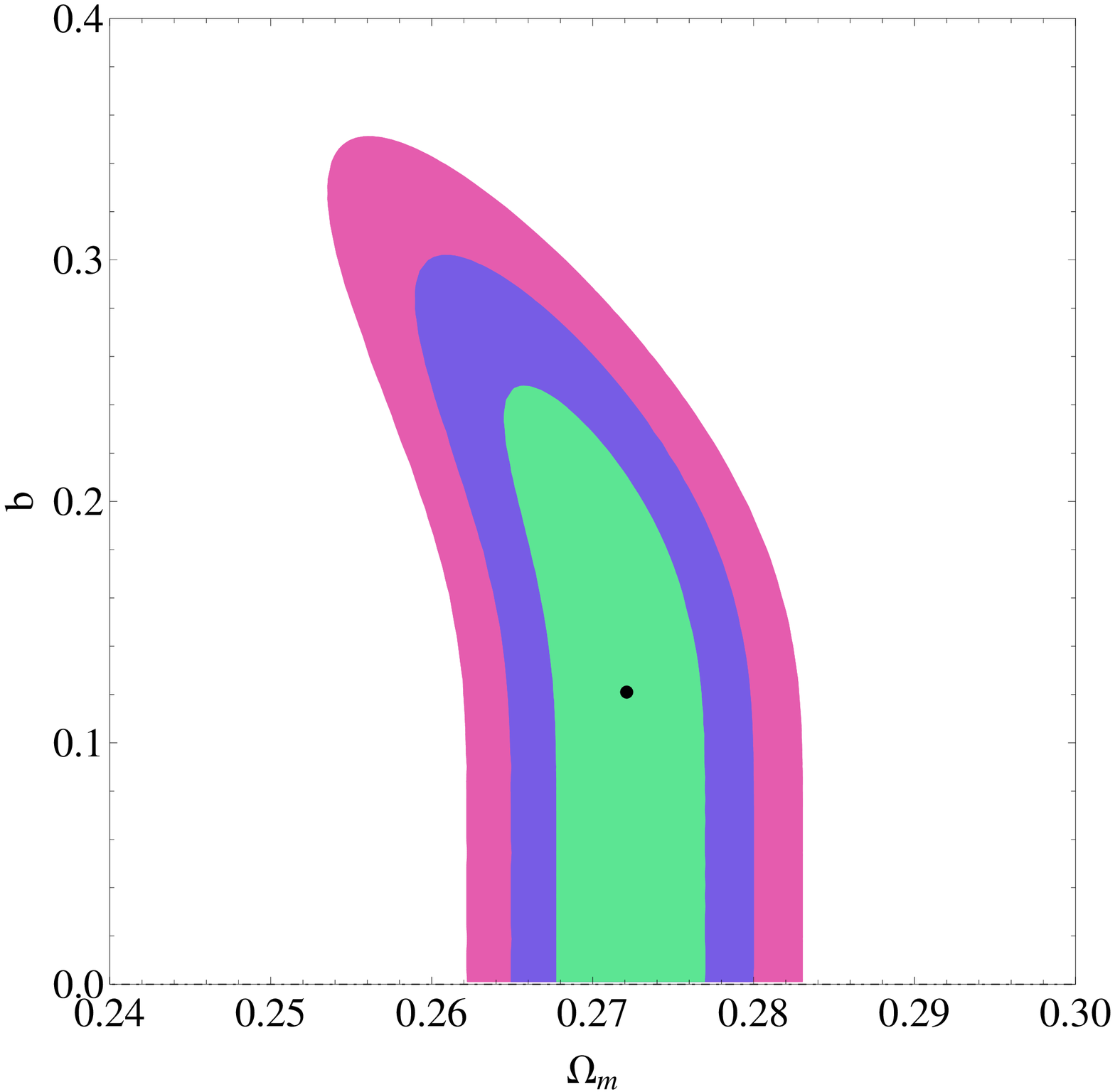}}}
\vspace{0cm}\rotatebox{0}{\vspace{0cm}\hspace{0cm}\resizebox{0.315\textwidth}{!}{\includegraphics{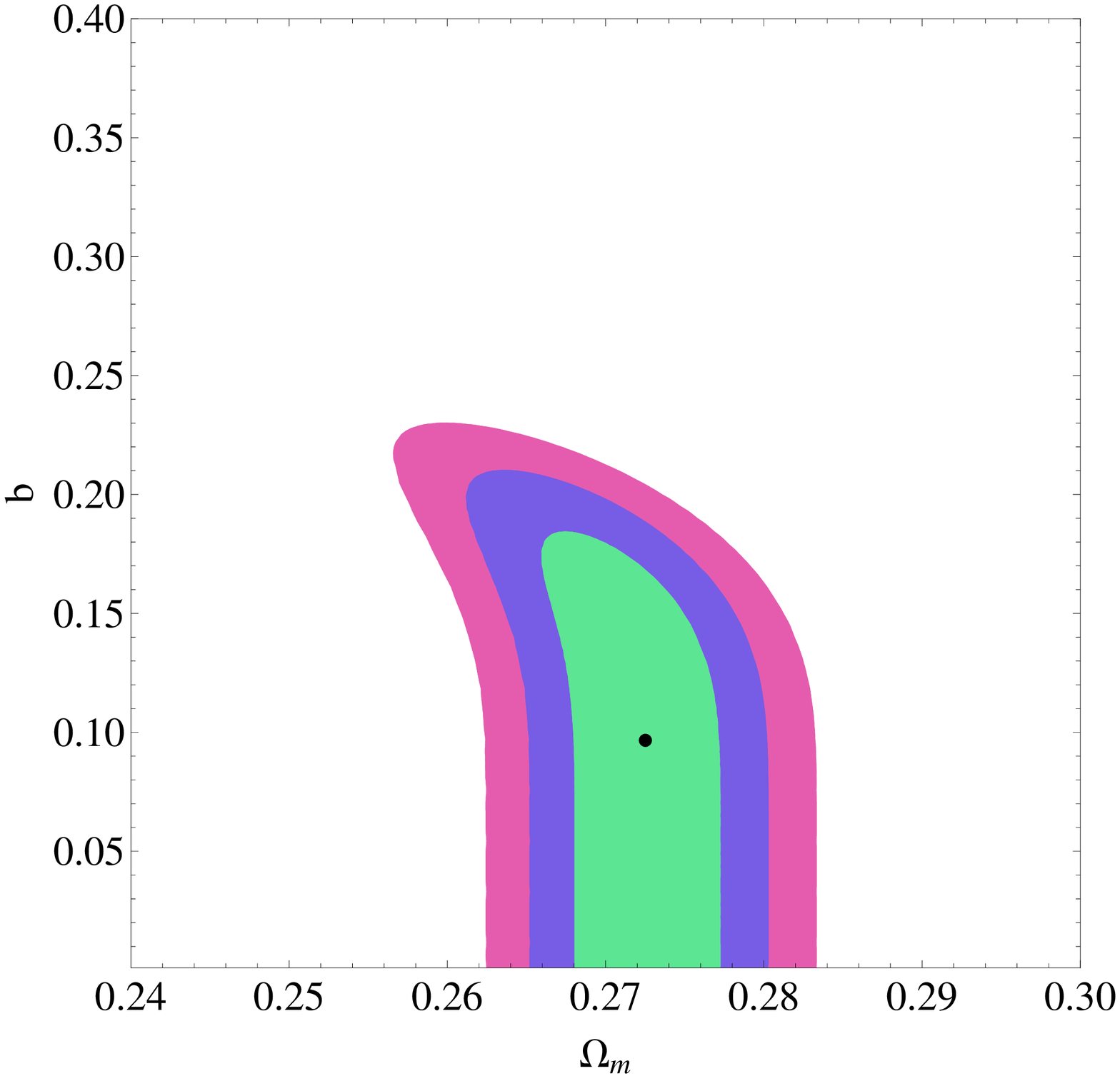}}}
\caption{Likelihood contours for $\delta \chi^{2}\equiv \chi^2-\chi^2_{min}$
equal to 2.30, 6.18 and 11.83, corresponding to 1$\sigma$, 2$\sigma$ and
$3\sigma$ confidence levels, in the $(\Omega_{m},b)$ plane for the $\Gamma_0$
growth rate parametrization and the $f_{1}$CDM (left),  $f_{2}$CDM (middle)
and $f_{3}$CDM (right) models. In all cases the black point corresponds to
the best fit. In this plot and in the ones that follow we have set the
parameters that are not shown to their best fit values for the corresponding
model (see Table \ref{tab:growth1}). \label{contoursomb}}
\end{figure*}

\begin{figure*}[!]
\centering
\vspace{0cm}\rotatebox{0}{\vspace{0cm}\hspace{0cm}\resizebox{0.31\textwidth}{!}{\includegraphics{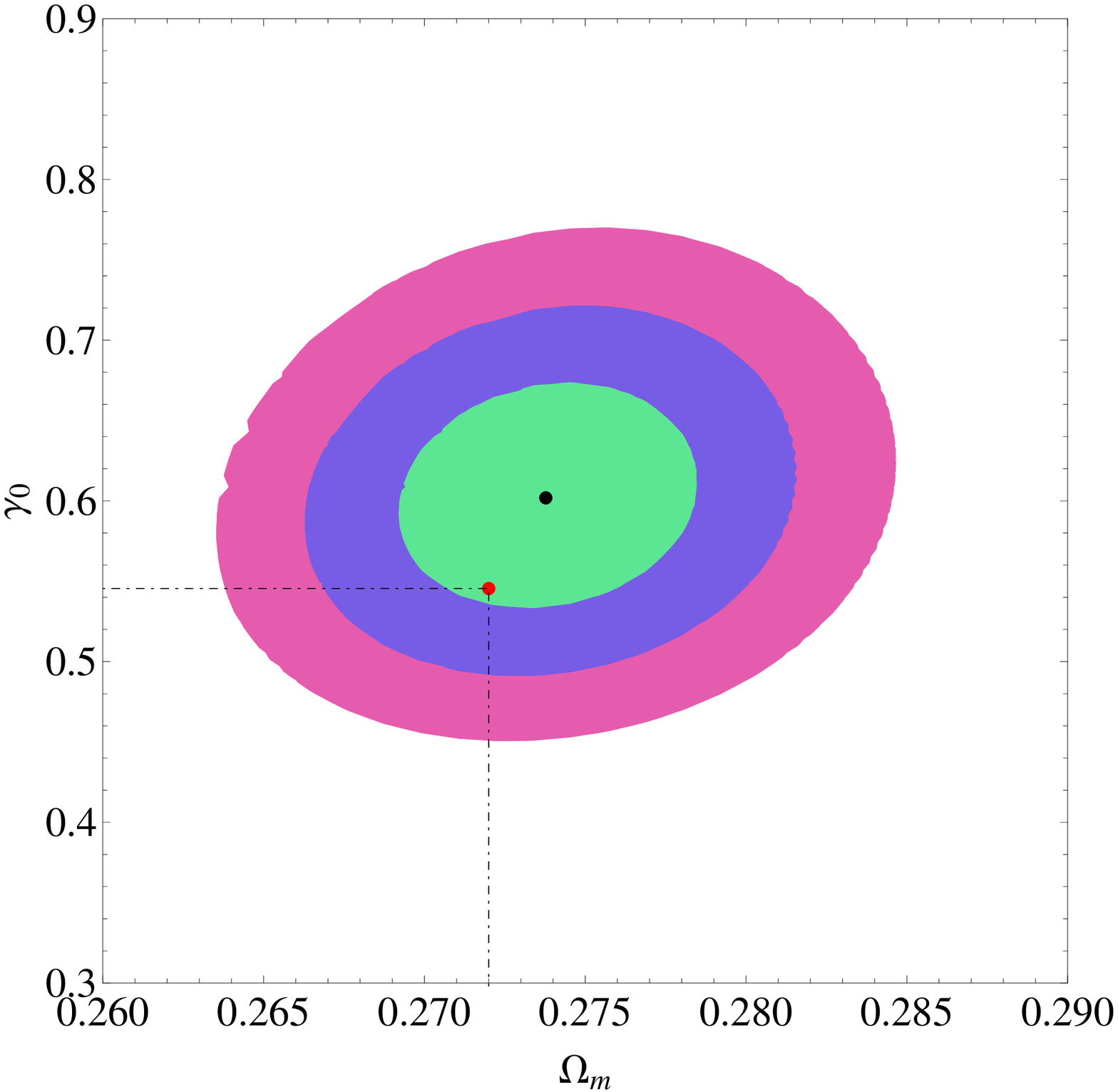}}}
\vspace{0cm}\rotatebox{0}{\vspace{0cm}\hspace{0cm}\resizebox{0.31\textwidth}{!}{\includegraphics{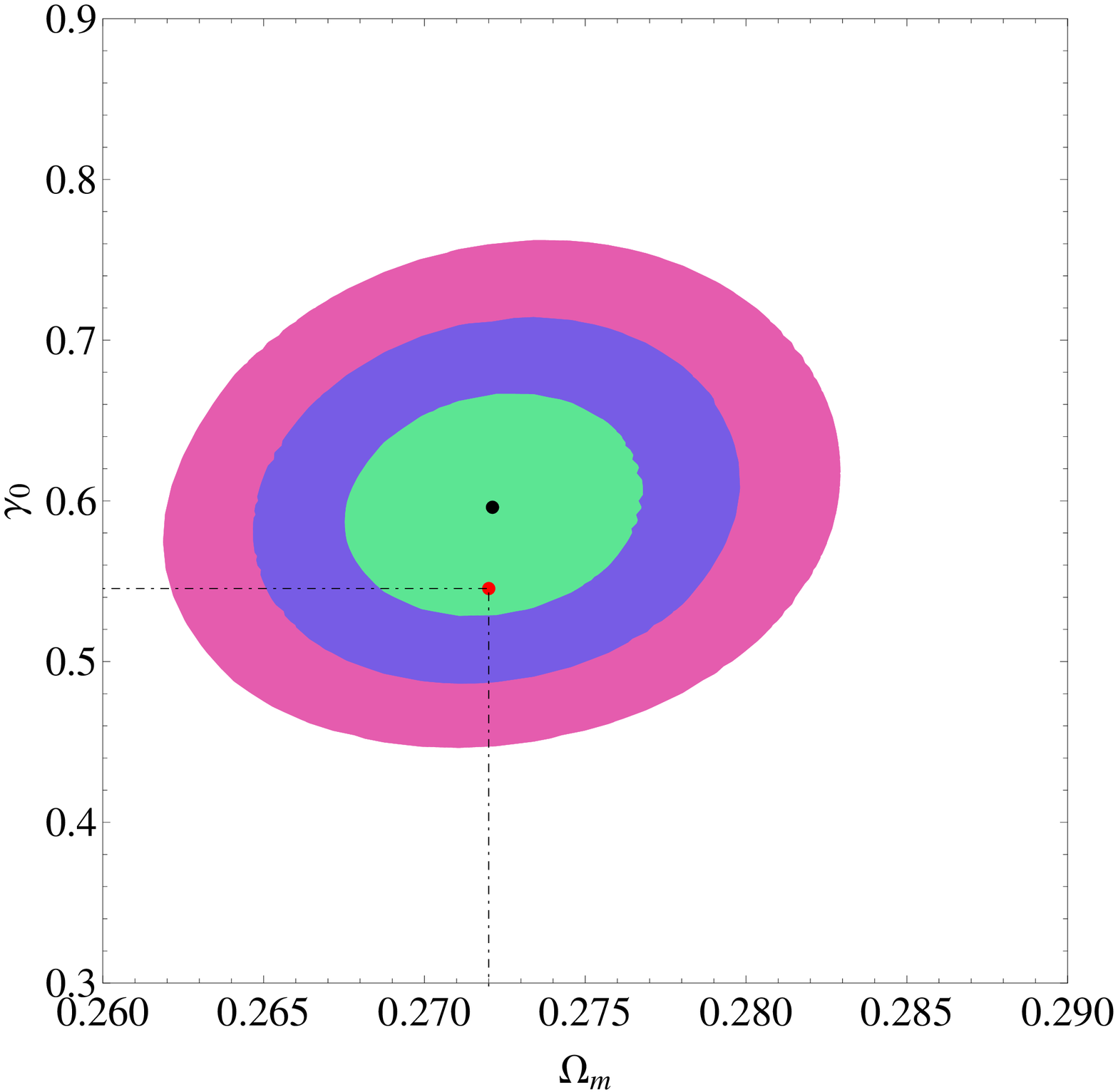}}}
\vspace{0cm}\rotatebox{0}{\vspace{0cm}\hspace{0cm}\resizebox{0.31\textwidth}{!}{\includegraphics{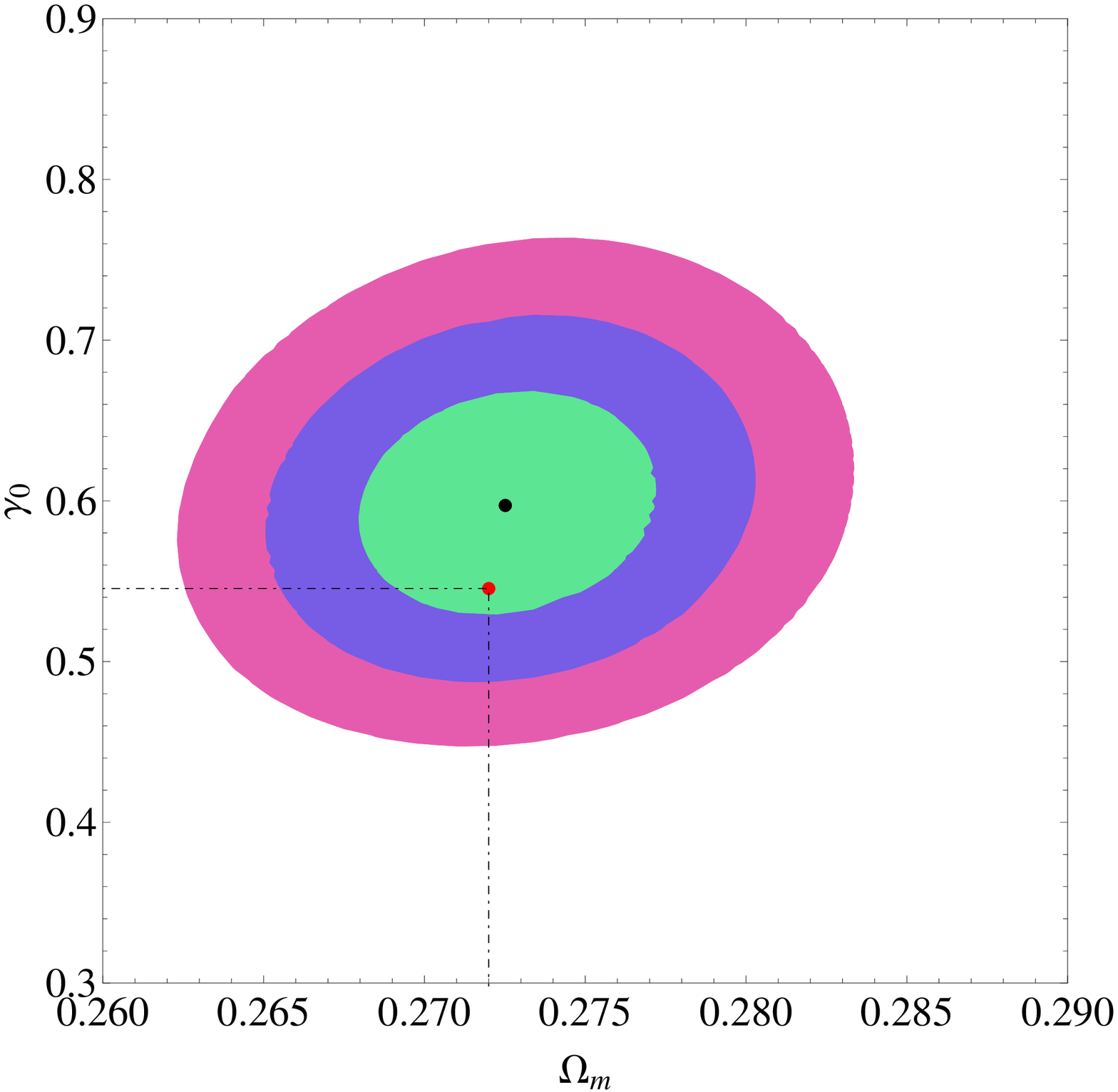}}}
\caption{Likelihood contours for $\delta \chi^{2}\equiv \chi^2-\chi^2_{min}$
equal to 2.30, 6.18 and 11.83, corresponding to 1$\sigma$, 2$\sigma$ and
$3\sigma$ confidence levels, in the $(\Omega_{m},\gamma)$ plane for the
$\Gamma_0$ growth rate parametrization and the $f_{1}$CDM (left), $f_{2}$CDM
(middle) and $f_{3}$CDM (right) models. In all cases the red point
corresponds to $(\Omega_{m},\gamma)=(0.272,6/11)$. \label{contoursG0}}
\end{figure*}

\noindent
(a) $\Gamma_{0}$ parametrization. -\vspace{0.1cm}\\
Regarding the $\Lambda$CDM cosmological model
our best fit value growth is $\gamma=0.597\pm 0.046$ that is  in a good
agreement with previous studies
\cite{Sam11,Bass,Por,Hud12,Samnew12,BasNes13}. Concerning the $f(T)$
models we obtain $(\gamma,b)=(0.602\pm 0.052,-0.017\pm 0.083)$,
$(\gamma,b)=(0.596\pm 0.047,0.121\pm 0.184)$ and $(\gamma,b)=(0.597\pm
0.046,0.097\pm 0.155)$ for the $f_1$CDM, $f_2$CDM and $f_3$CDM models,
respectively, with a reduced $\chi^{2}_{min}$ of $\sim 574.2$.
In Fig.~\ref{contoursomb} we show the 1$\sigma$, 2$\sigma$ and $3\sigma$
confidence contours in the $(\Omega_{m},b)$ plane, while in
Fig.~\ref{contoursG0} we present the corresponding contours in the
$(\Omega_{m},\gamma)$ plane.

\begin{figure*}[!]
\centering
\vspace{0cm}\rotatebox{0}{\vspace{0cm}\hspace{0cm}\resizebox{0.31\textwidth}{!}{\includegraphics{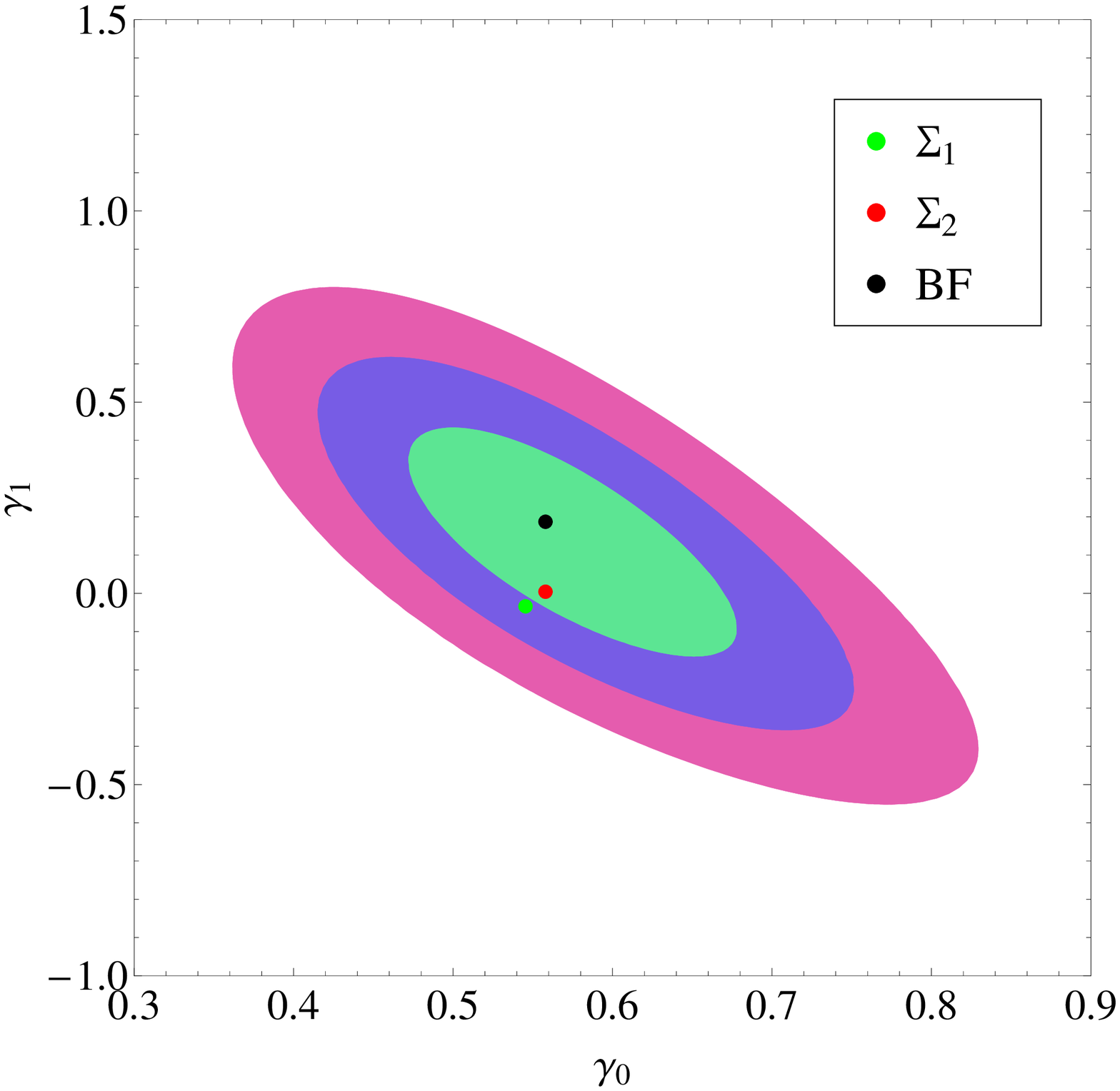}}}
\vspace{0cm}\rotatebox{0}{\vspace{0cm}\hspace{0cm}\resizebox{0.31\textwidth}{!}{\includegraphics{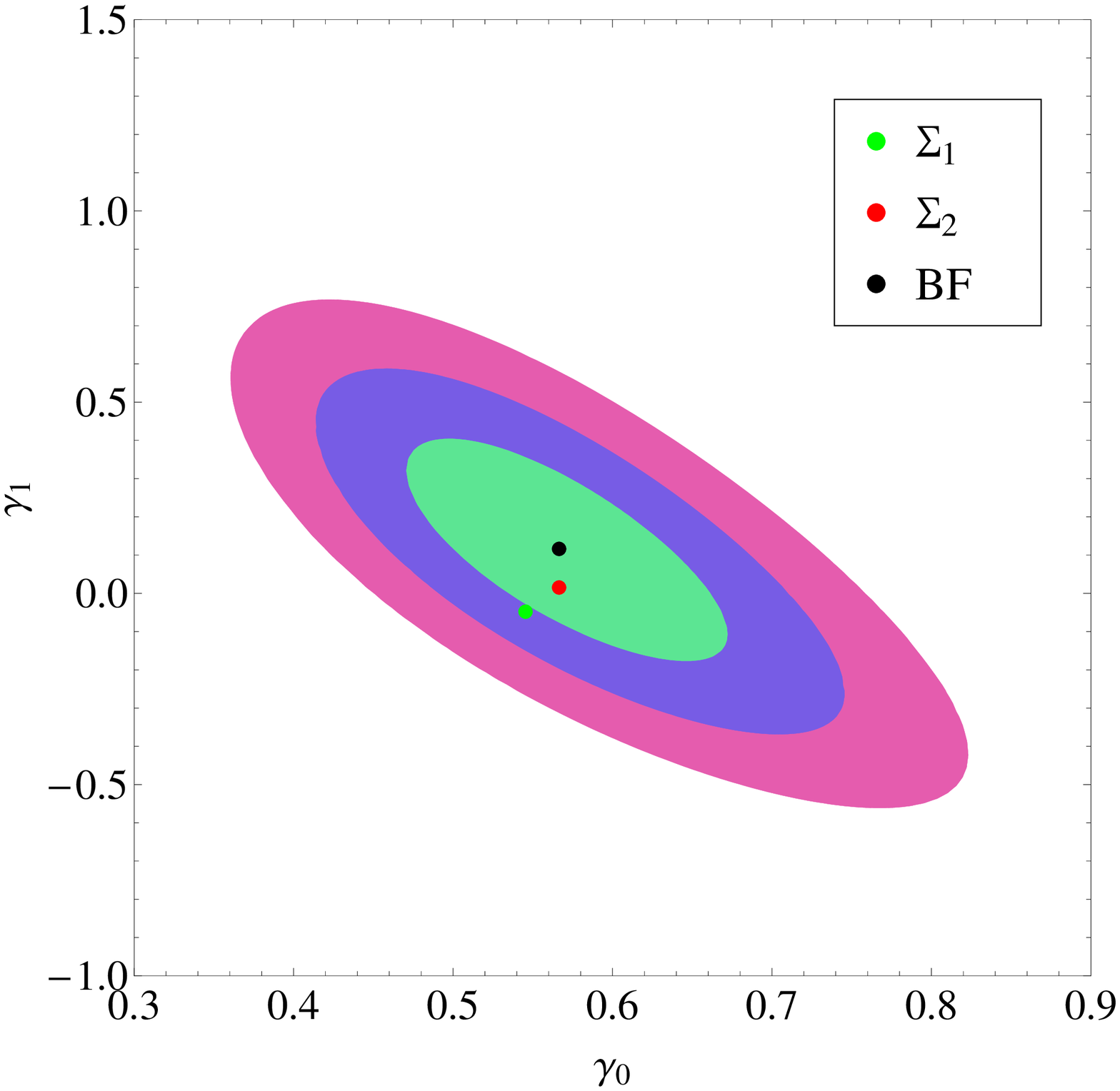}}}
\vspace{0cm}\rotatebox{0}{\vspace{0cm}\hspace{0cm}\resizebox{0.31\textwidth}{!}{\includegraphics{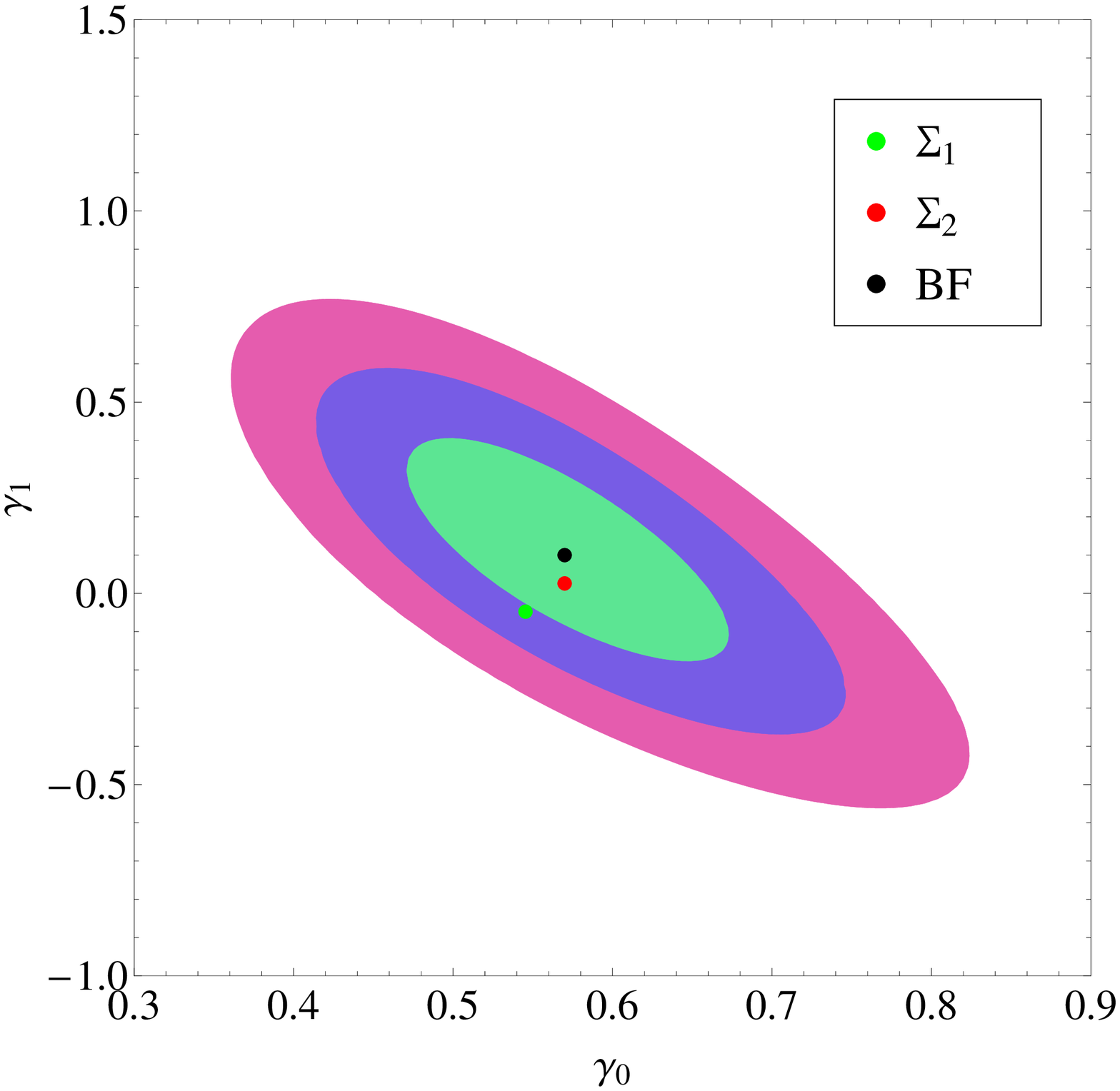}}}
\caption{Likelihood contours for $\delta \chi^{2}\equiv \chi^2-\chi^2_{min}$
equal to 2.30, 6.18 and 11.83, corresponding to 1$\sigma$, 2$\sigma$ and
$3\sigma$ confidence levels, in the $(\gamma_{0},\gamma_{1})$ plane for the
$\Gamma_1$ growth rate parametrization and for the $f_{1}$CDM (left),
$f_{2}$CDM (middle) and $f_{3}$CDM (right) models. We also include the
theoretical $\Lambda$CDM $(\gamma_{0},\gamma_{1})$ values given by
$\Sigma_1=\left(6/11,\gamma_1(6/11,\Omega_{m0,bf})\right)$ and
$\Sigma_2=\left(\gamma_{0,bf},\gamma_1(\gamma_{0,bf},\Omega_{m0,bf})\right)$.
\label{contoursG1}}
\end{figure*}

\noindent (b) $\Gamma_{1}$ parametrization. - \vspace{0.1cm}\\
In the case of the concordance $\Lambda$ cosmology we find
$\gamma_{0}=0.567\pm  0.066$ and $\gamma_{1}=0.116\pm 0.191$
with $\chi_{min}^{2} \simeq 573.861$ which are in agreement
with previous studies \cite{Nes08,Port08,Dos10,Por,BasNes13}.
For the $f_{1}$CDM, $f_{2}$CDM and $f_{3}$CDM models the corresponding
likelihood functions peak at $(b,\gamma_{0},\gamma_{1})=(-0.029\pm 0.088,0.558\pm 0.067,0.187\pm 0.205$) with $\chi_{min}^{2} \simeq 573.817$,
$(b,\gamma_{0},\gamma_{1})=(0.086\pm 0.301,0.566\pm 0.066,0.116\pm 0.191$)
with $\chi_{min}^{2} \simeq 573.863$ and $(b,\gamma_{0},\gamma_{1})=(0.010\pm 0.324,0.570\pm 0.067,0.099\pm 0.192$) with $\chi_{min}^{2} \simeq 573.852$, respectively. In Fig.~\ref{contoursG1} we present the corresponding 1$\sigma$, 2$\sigma$ and $3\sigma$ contours in the $(\gamma_{0},\gamma_{1})$ plane.

\begin{figure*}[!]
\centering
\vspace{0cm}\rotatebox{0}{\vspace{0cm}\hspace{0cm}\resizebox{0.31\textwidth}{!}{\includegraphics{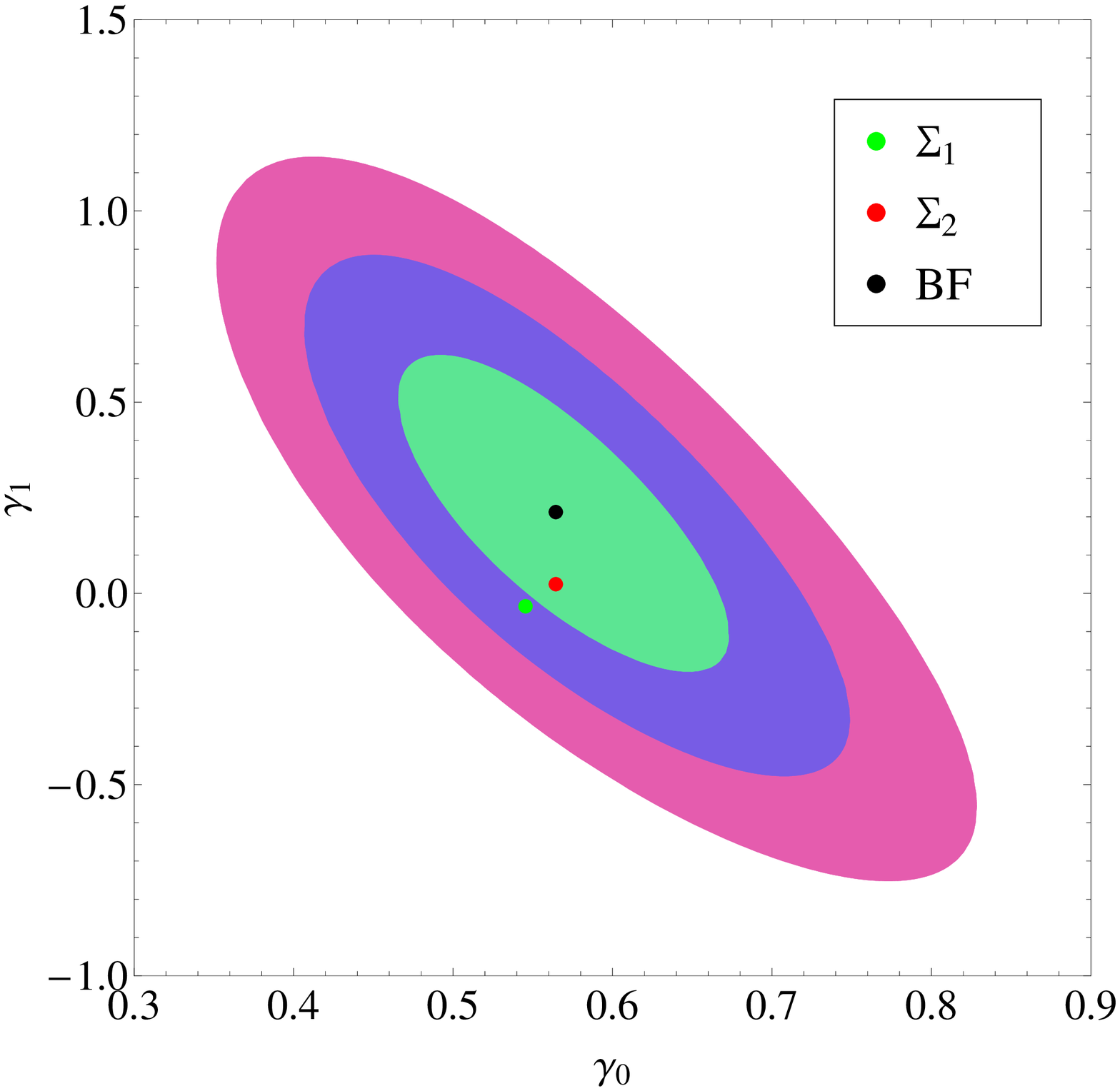}}}
\vspace{0cm}\rotatebox{0}{\vspace{0cm}\hspace{0cm}\resizebox{0.31\textwidth}{!}{\includegraphics{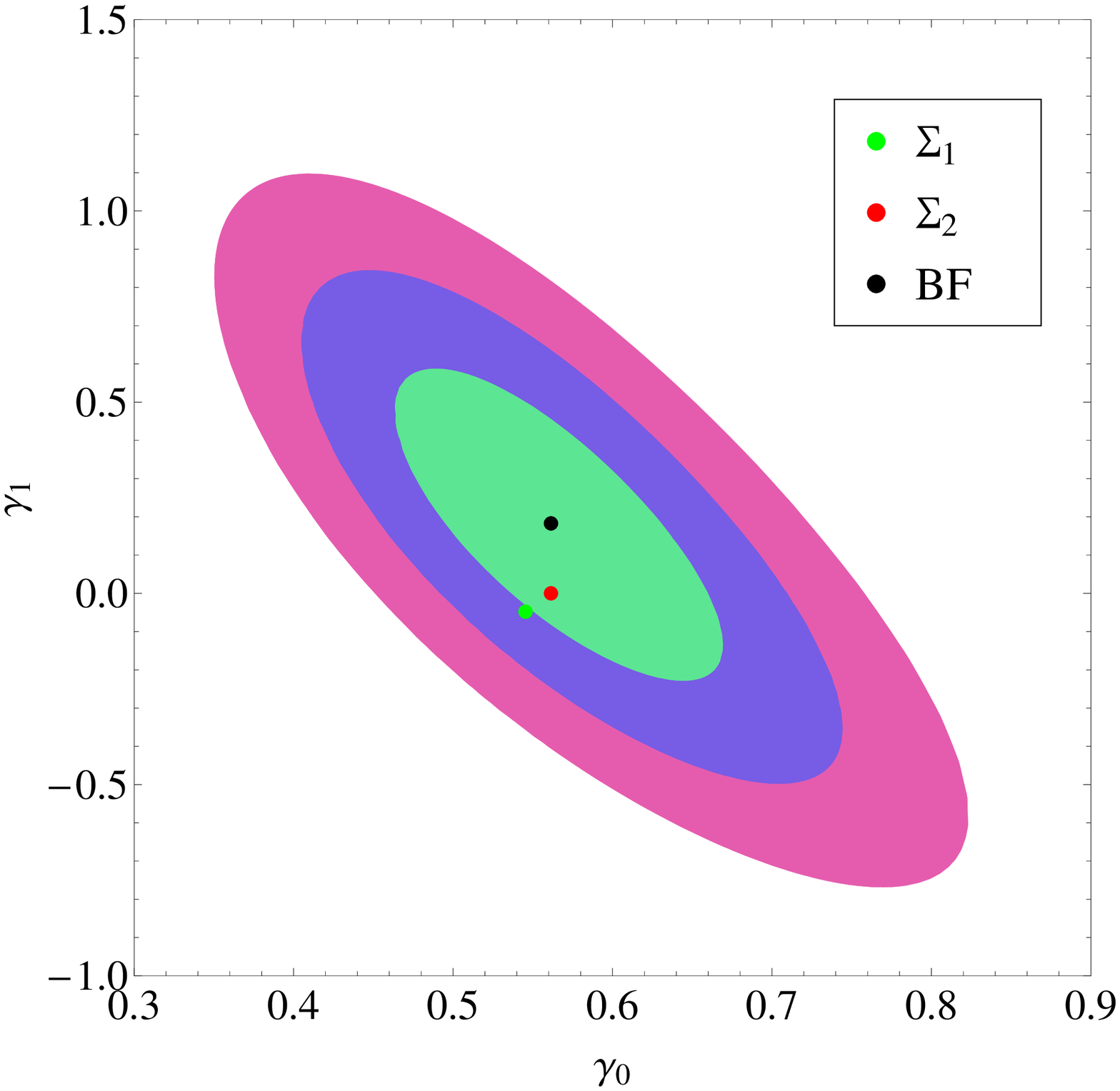}}}
\vspace{0cm}\rotatebox{0}{\vspace{0cm}\hspace{0cm}\resizebox{0.31\textwidth}{!}{\includegraphics{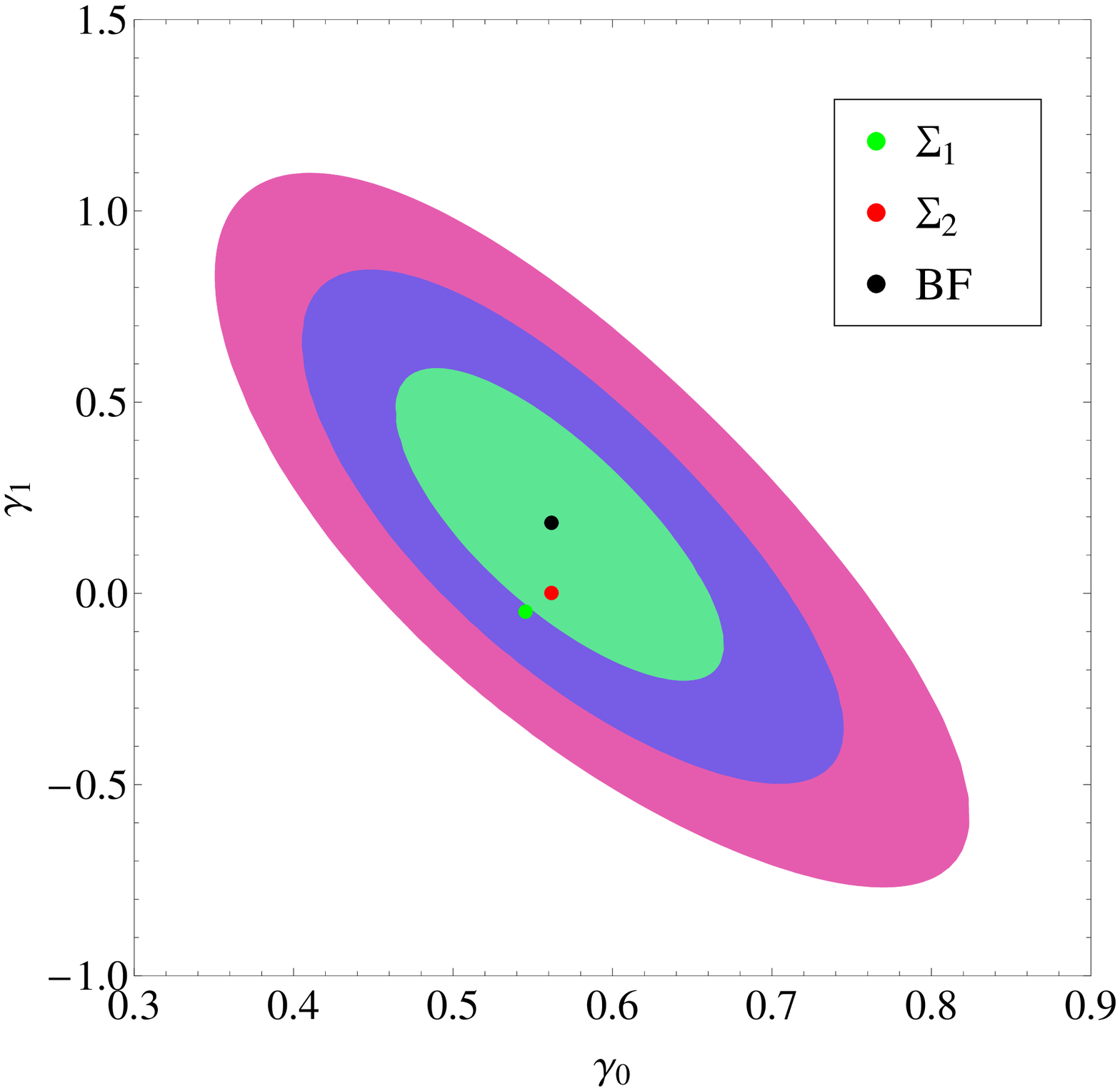}}}
\caption{Likelihood contours for $\delta \chi^{2}\equiv \chi^2-\chi^2_{min}$
equal to 2.30, 6.18 and 11.83, corresponding to 1$\sigma$, 2$\sigma$ and
$3\sigma$ confidence levels, in the $(\gamma_{0},\gamma_{1})$ plane for the
$\Gamma_2$ growth rate parametrization and for the $f_{1}$CDM (left),
$f_{2}$CDM (middle) and $f_{3}$CDM (right) models. We also include the
theoretical $\Lambda$CDM $(\gamma_{0},\gamma_{1})$ values given by
$\Sigma_1=\left(6/11,\gamma_1(6/11,\Omega_{m0,bf})\right)$ and
$\Sigma_2=\left(\gamma_{0,bf},\gamma_1(\gamma_{0,bf},\Omega_{m0,bf})\right)$.
\label{contoursG2}}
\end{figure*}

\noindent
(c) $\Gamma_{2}$ parametrization. - \vspace{0.1cm}\\
In the case of
$\Lambda$CDM model we have
$\gamma_{0}=0.561\pm0.068$,
$\gamma_{1}=0.183\pm0.269$ ($\chi_{min}^{2} \simeq 573.767$), while for
the $f_{1}$CDM we obtain $b=-0.030\pm 0.089$,
$\gamma_{0}=0.564\pm0.069$,
$\gamma_{1}=0.213\pm0.287$ ($\chi_{min}^{2} \simeq 573.640$),
for the $f_{2}$CDM gravity model
we find $b=0.150\pm 0.096$,
$\gamma_{0}=0.560\pm0.068$,
$\gamma_{1}=0.181\pm0.271$ ($\chi_{min}^{2} \simeq 573.921$) and finally for the $f_{3}$CDM model we have
we find $b=0.024\pm 0.183$,
$\gamma_{0}=0.562\pm0.068$,
$\gamma_{1}=0.185\pm0.269$ ($\chi_{min}^{2} \simeq 573.749$).
In Fig.~\ref{contoursG2} we present the corresponding 1$\sigma$, 2$\sigma$ and $3\sigma$ contours in the $(\gamma_{0},\gamma_{1})$ plane.

We stress here that in all three previous $f(T)$ models, namely,
$f_{1-3}$CDM ones, the parameter $b$ which quantifies the deviation
from $\Lambda$CDM cosmology is constrained in a very narrow window around
$0$. Thus, although these three models are consistent with observations,
their viable forms are practically  indistinguishable from $\Lambda$CDM and
therefore their new degrees of freedom are disfavored by data.

\begin{figure*}[!]
\centering
\vspace{0cm}\rotatebox{0}{\vspace{0cm}\hspace{0cm}\resizebox{0.3\textwidth}{!}{\includegraphics{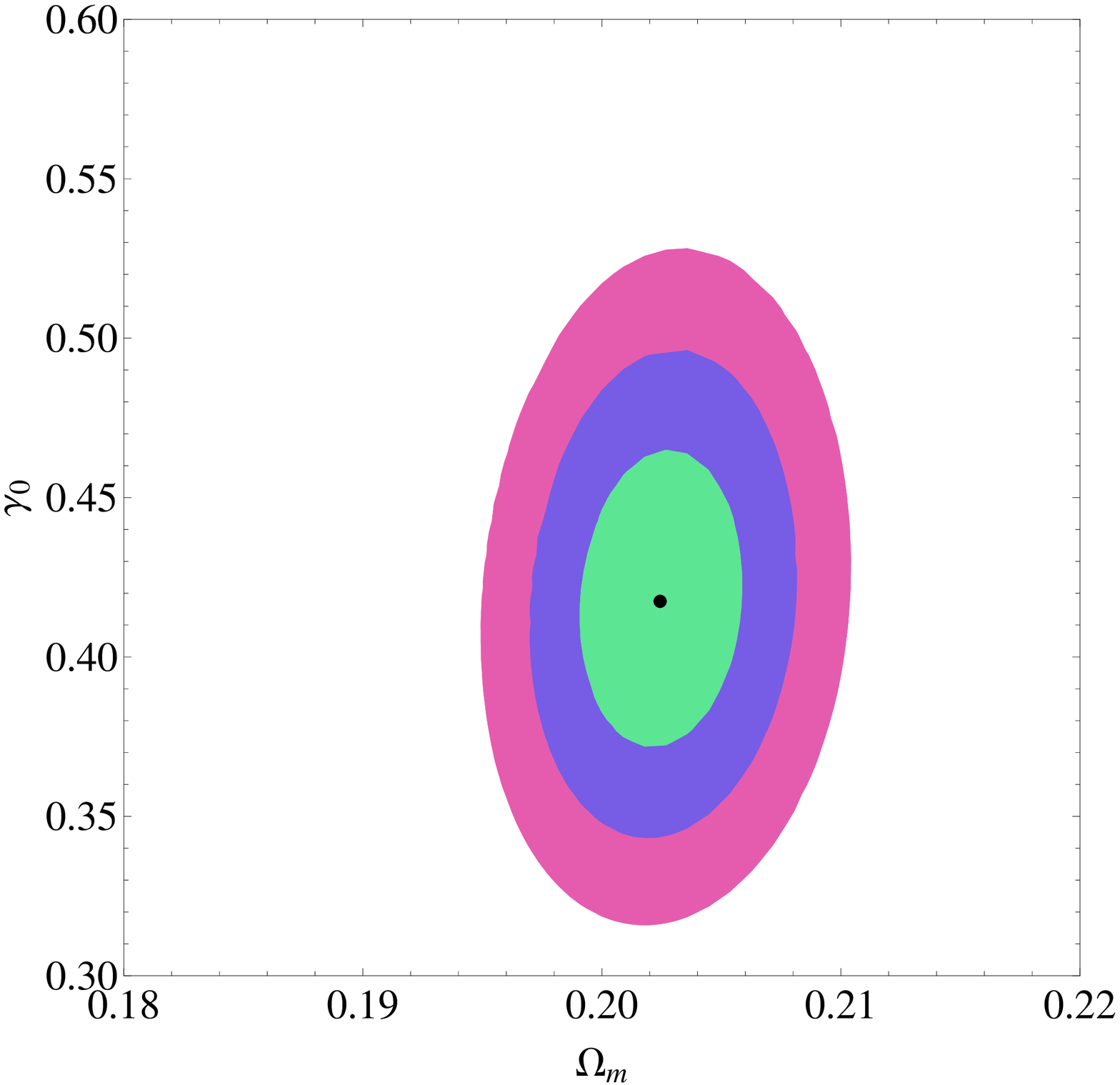}}}
\vspace{0cm}\rotatebox{0}{\vspace{0cm}\hspace{0cm}\resizebox{0.3\textwidth}{!}{\includegraphics{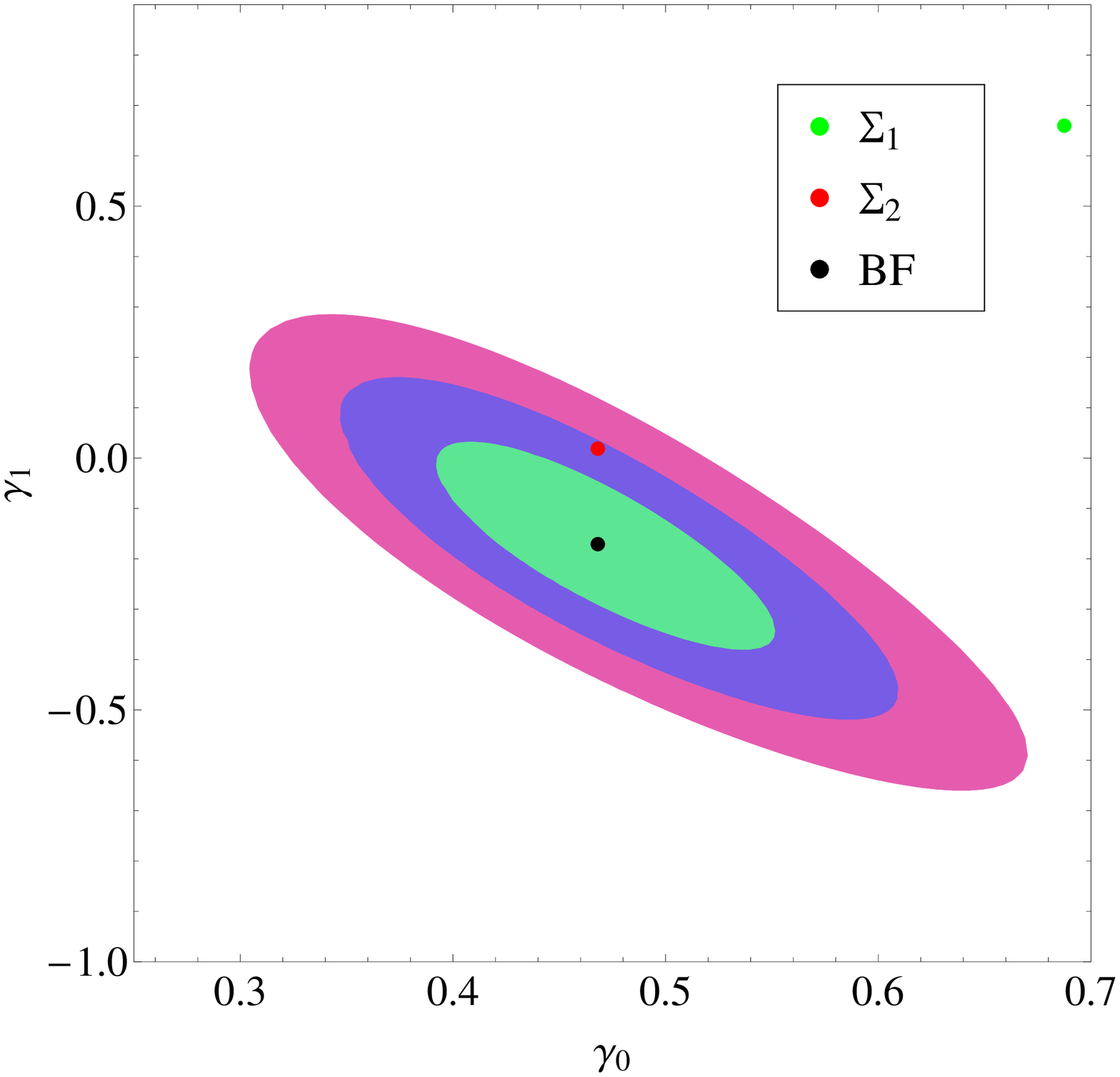}}}
\vspace{0cm}\rotatebox{0}{\vspace{0cm}\hspace{0cm}\resizebox{0.3\textwidth}{!}{\includegraphics{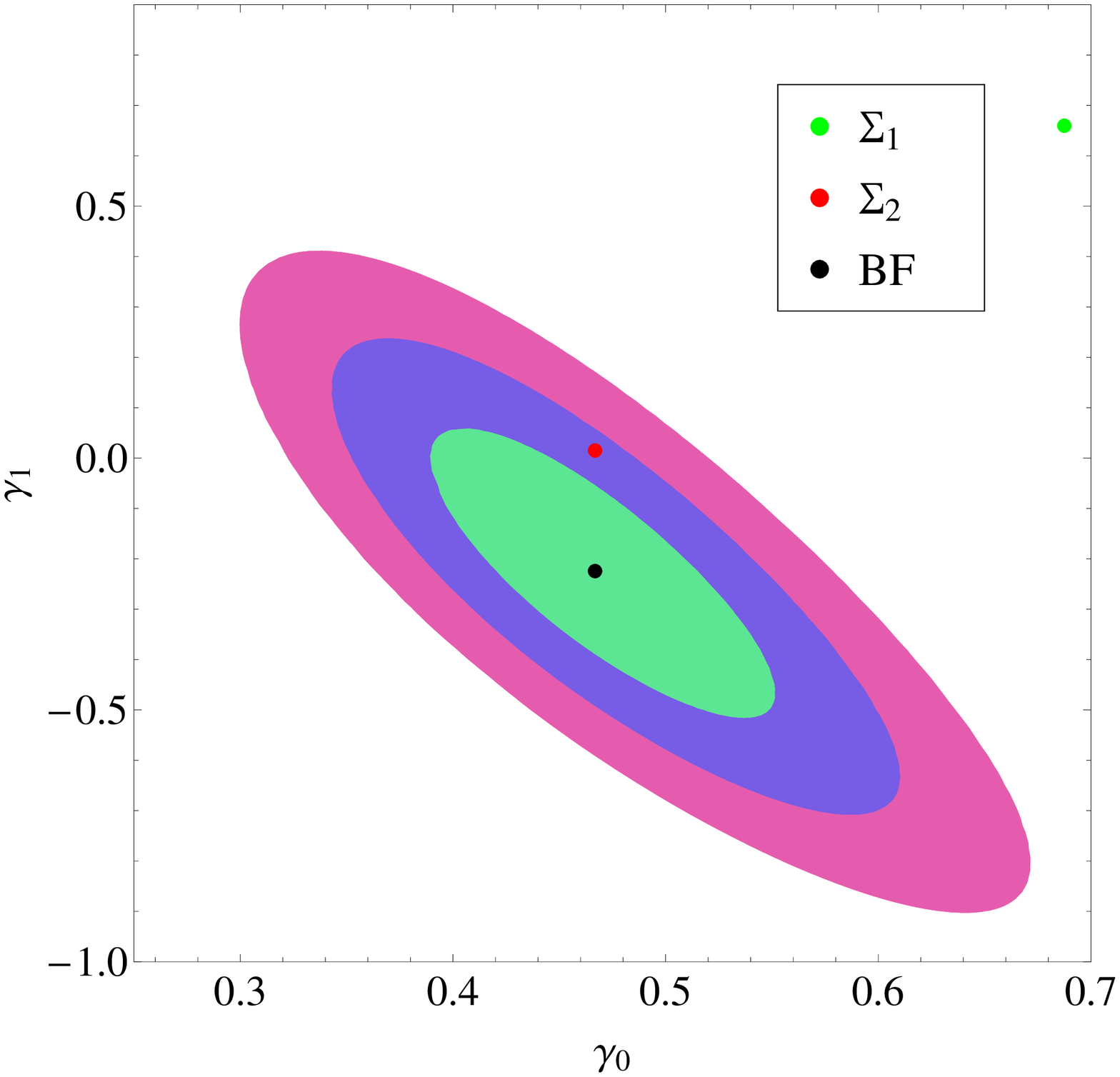}}}
\caption{Likelihood contours for $\delta \chi^{2}\equiv \chi^2-\chi^2_{min}$
equal to 2.30, 6.18 and 11.83, corresponding to 1$\sigma$, 2$\sigma$ and
$3\sigma$ confidence levels, for the $f_{4}$CDM model in the
$(\Omega_m,\gamma_{0})$ plane (left) and the $(\gamma_{0},\gamma_{1})$ plane
(middle) and (right). We also include the theoretical $\Lambda$CDM
$(\gamma_{0},\gamma_{1})$ values given by
$\Sigma_1=\left(11/16,\gamma_1(11/16,\Omega_{m0,bf})\right)$ (with the value
$\gamma_0=11/16$ corresponding to the DGP) and
$\Sigma_2=\left(\gamma_{0,bf},\gamma_1(\gamma_{0,bf},\Omega_{m0,bf})\right)$.
As was mentioned in the text, the difference between the DGP (green point)
and $f_{4}$CDM (black point) is due to the different  $G_{\rm eff}(z)$, which
affects the evolution of the matter density perturbations.
\label{contoursf4}}
\end{figure*}

Finally, in Fig.~\ref{contoursf4} we show the likelihood contours for
$f_{4}$CDM model, which as discussed in Sec. \ref{fTmodels} coincides
with DGP at the background level, and thus it shares its observational
disadvantages and therefore we consider it as nonviable. In the same lines,
as we can see from Table \ref{tab:growth1}, for $f_{5}$CDM model we obtain
the best fits $\chi_{min}^2=(579.583, 580.723,
578.027)$ for the $\Gamma_0$, $\Gamma_1$ and $\Gamma_2$ growth-rate
parameterizations respectively, while it additionally has one more free
parameter than $\Lambda$CDM. Thus, this model is in tension with the
data.

\begin{figure*}[!]
\centering
\vspace{0cm}\rotatebox{0}{\vspace{0cm}\hspace{0cm}\resizebox{0.49\textwidth}{!}{\includegraphics{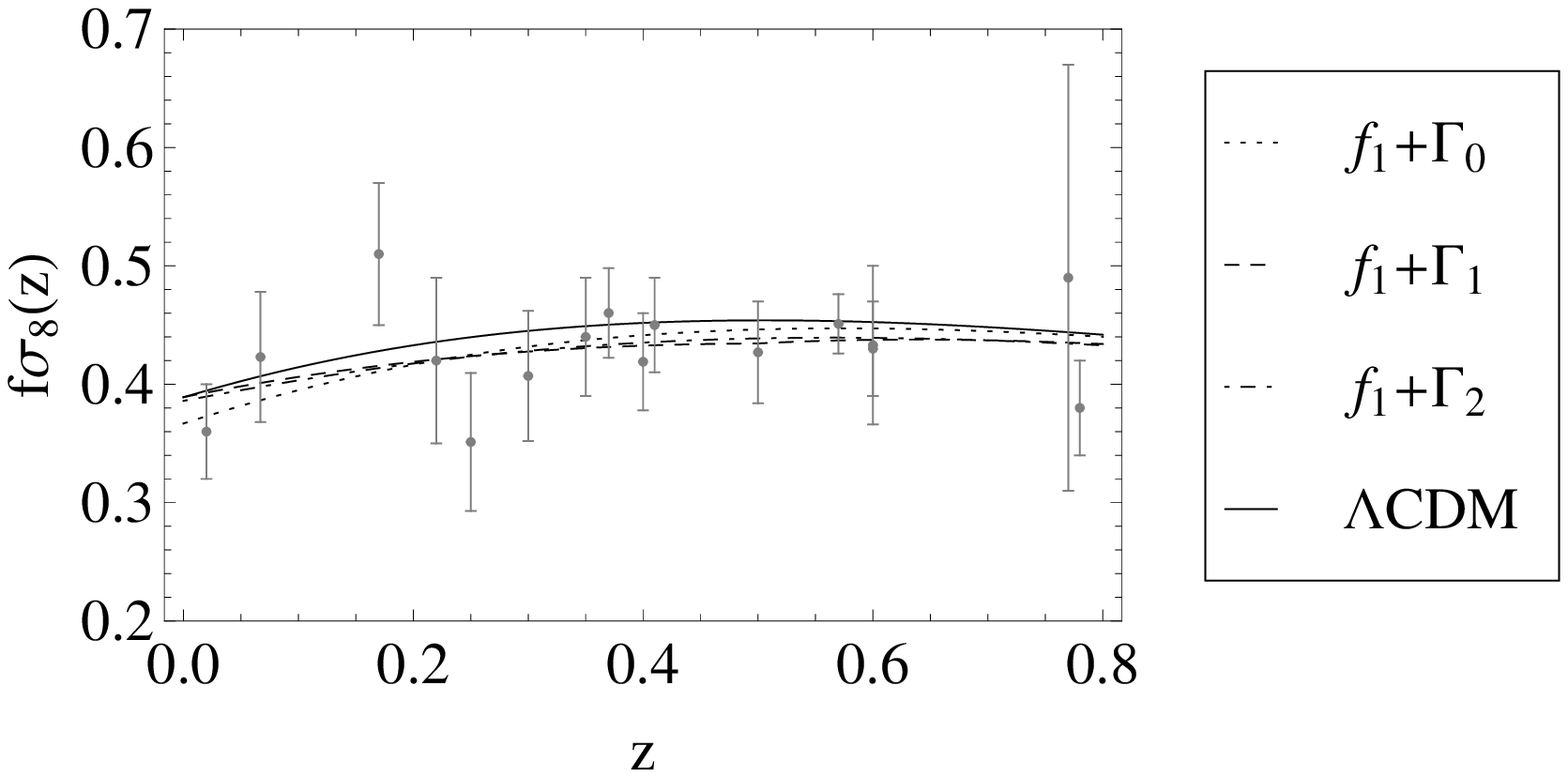}}}
\vspace{0cm}\rotatebox{0}{\vspace{0cm}\hspace{0cm}\resizebox{0.49\textwidth}{!}{\includegraphics{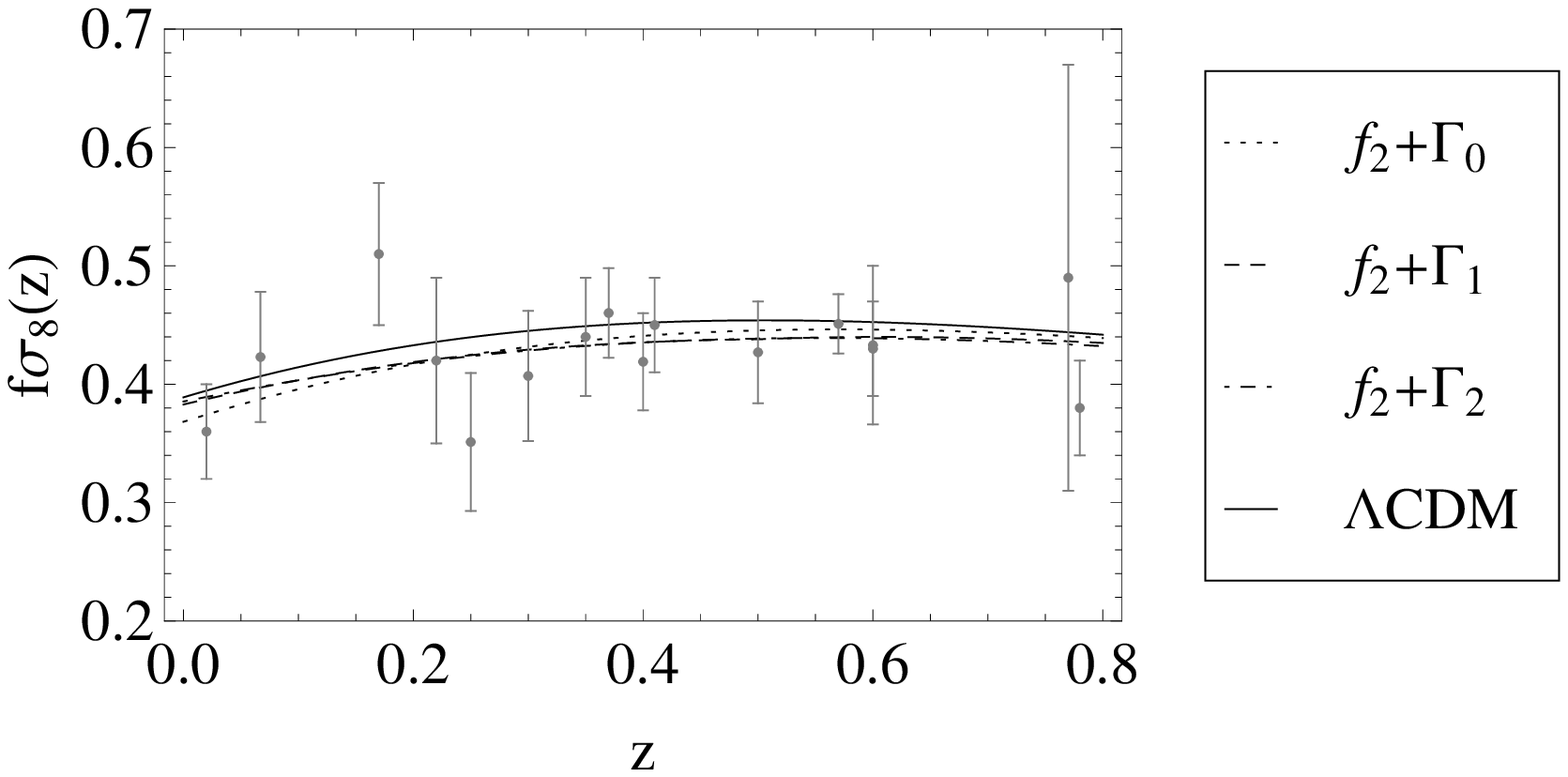}}}
\vspace{0cm}\rotatebox{0}{\vspace{0cm}\hspace{0cm}\resizebox{0.49\textwidth}{!}{\includegraphics{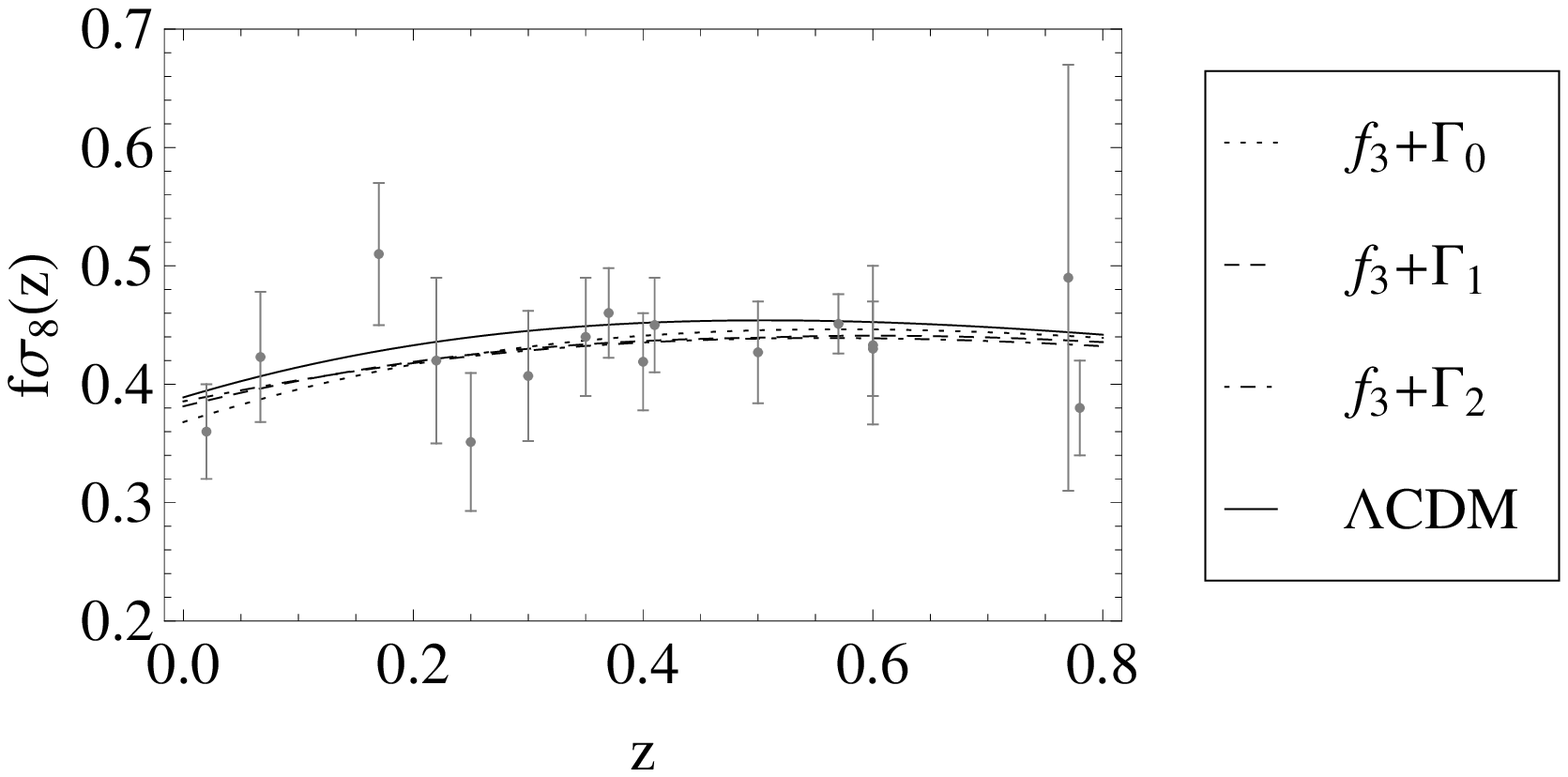}}}
\vspace{0cm}\rotatebox{0}{\vspace{0cm}\hspace{0cm}\resizebox{0.49\textwidth}{!}{\includegraphics{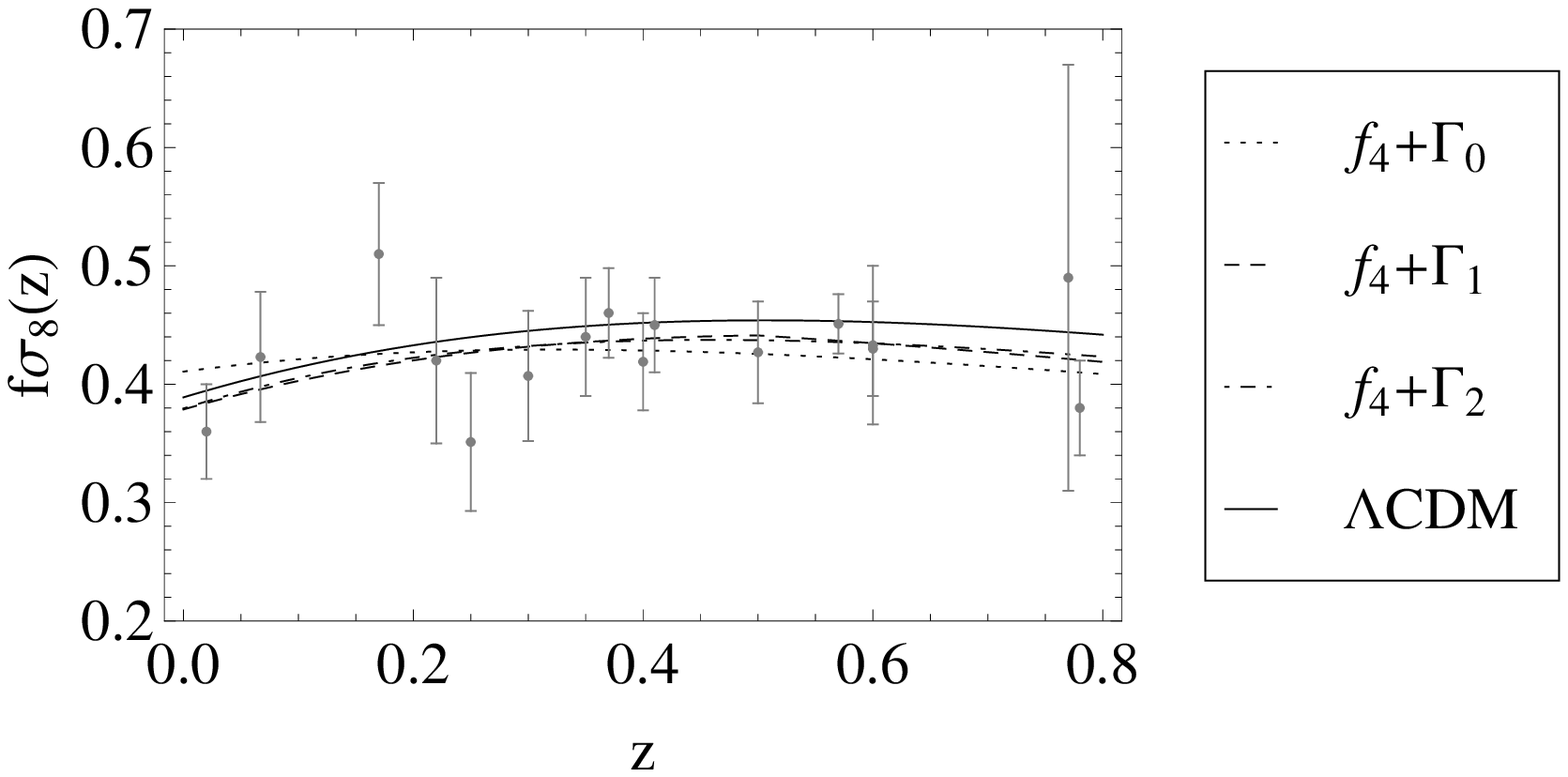}}}
\caption{Comparison of the observed and theoretical evolution of the growth rate $\fs(z)=F(z)\sigma_{8}(z)$ for the $f_{1-4}$CDM models [$f_{1}$CDM (top left), $f_{2}$CDM (top right), $f_{3}$CDM (bottom left), $f_{4}$CDM (bottom right)] and the various growth rate parameterizations. The dotted, dashed and dot-dashed lines correspond to the best fit $\Gamma_0$, $\Gamma_1$ and $\Gamma_2$ parametrizations while the solid black line corresponds to the exact solution of Eq.~(\ref{odedelta}) for $\fs(z)$ for the $\Lambda$CDM model for $\Omega_{m}=0.273$
\cite{Hinshaw:2012fq}. \label{growthrate}}
\end{figure*}

\begin{figure*}[!]
\centering
\vspace{0cm}\rotatebox{0}{\vspace{0cm}\hspace{0cm}\resizebox{0.45\textwidth}{!}{\includegraphics{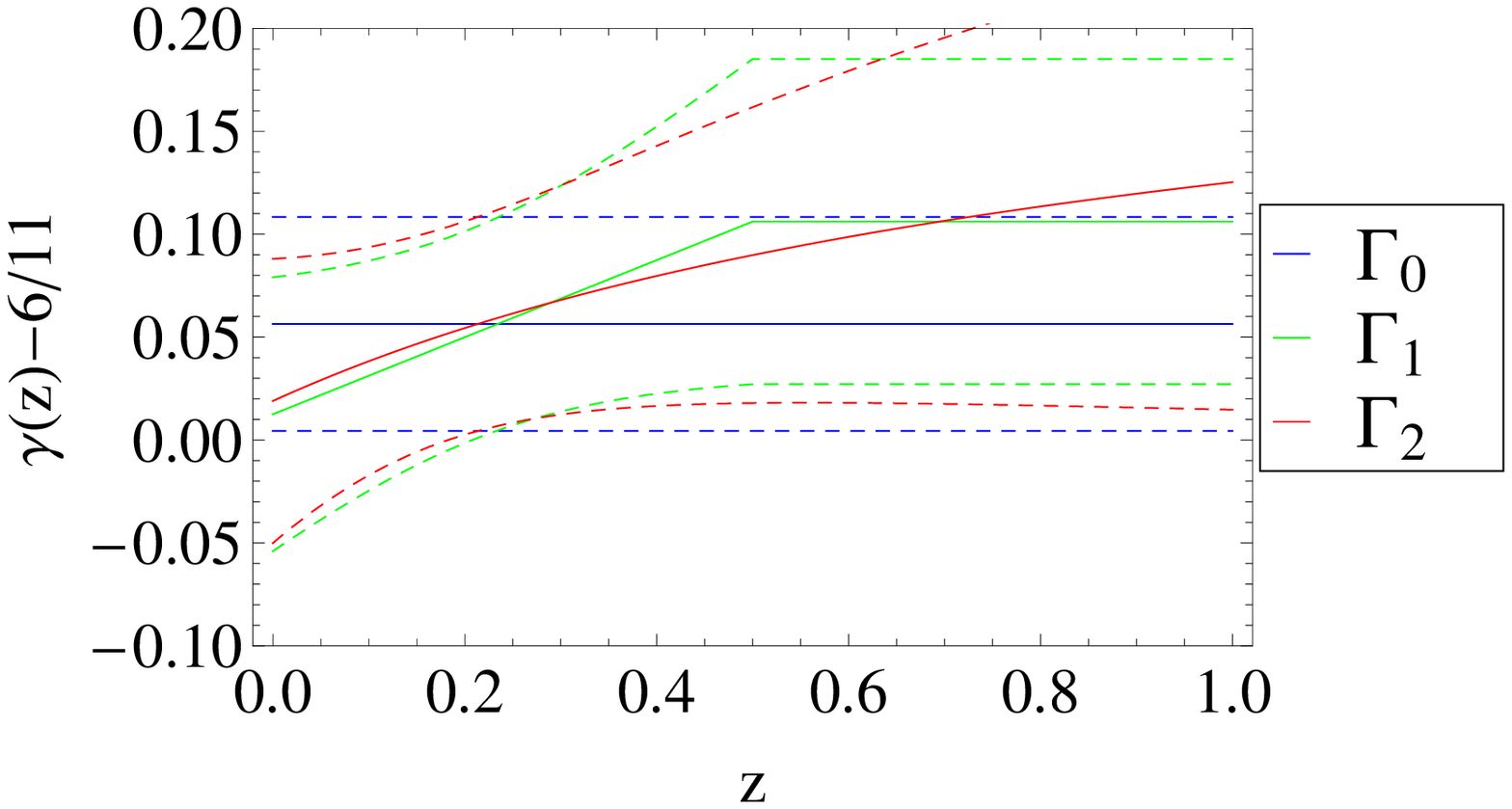}}}
\vspace{0cm}\rotatebox{0}{\vspace{0cm}\hspace{0cm}\resizebox{0.45\textwidth}{!}{\includegraphics{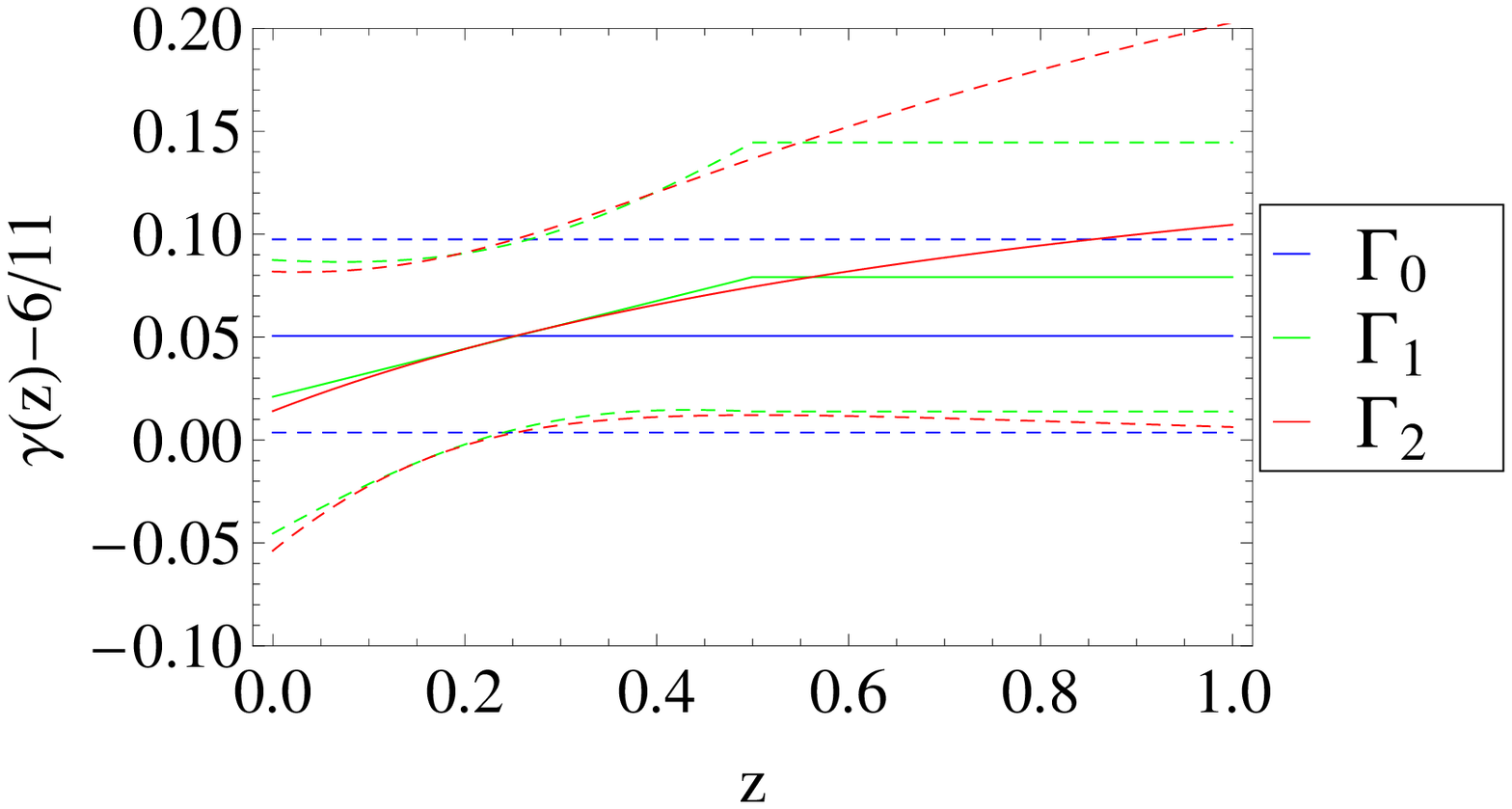}}}
\vspace{0cm}\rotatebox{0}{\vspace{0cm}\hspace{0cm}\resizebox{0.45\textwidth}{!}{\includegraphics{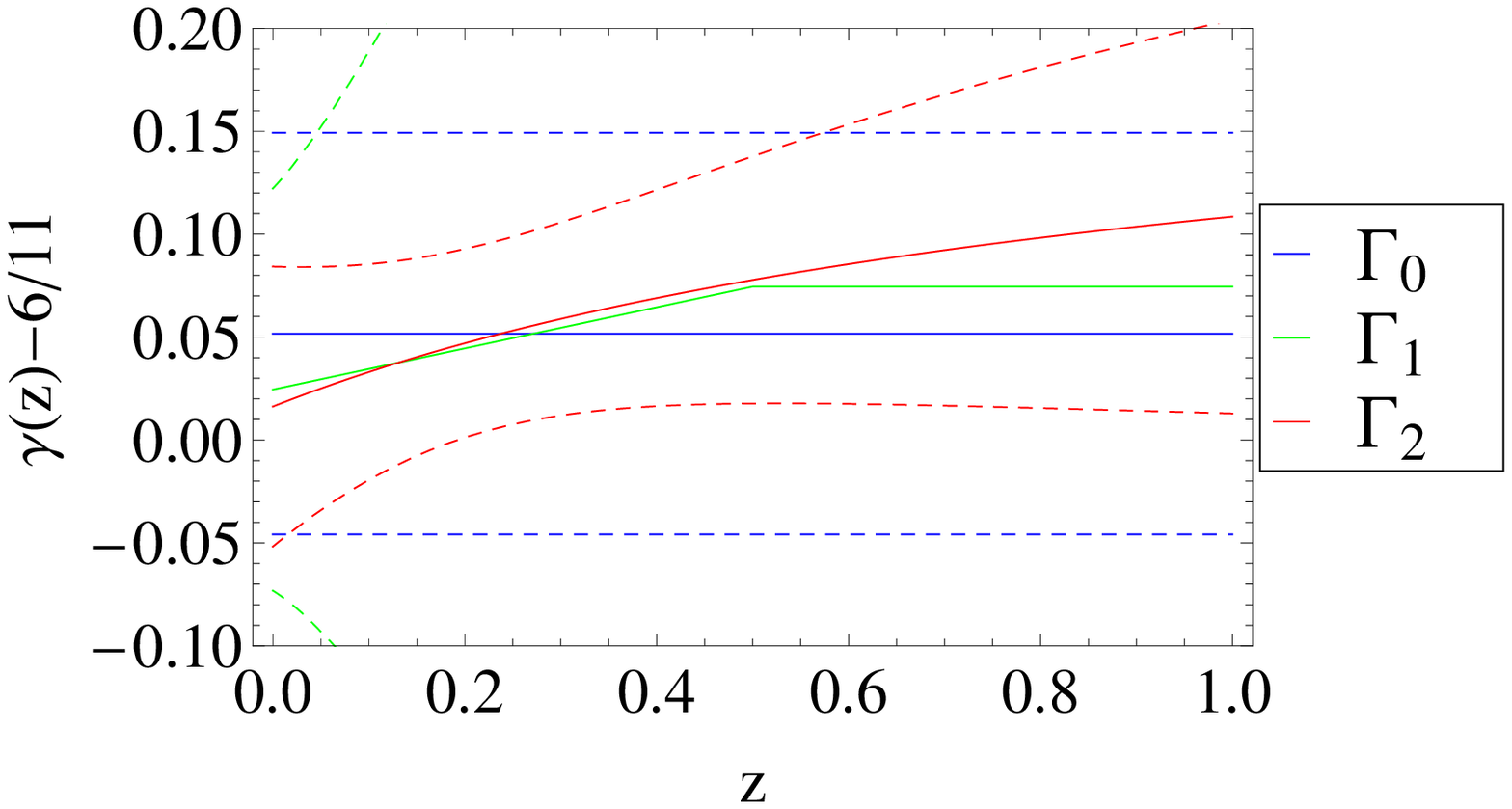}}}
\vspace{0cm}\rotatebox{0}{\vspace{0cm}\hspace{0cm}\resizebox{0.45\textwidth}{!}{\includegraphics{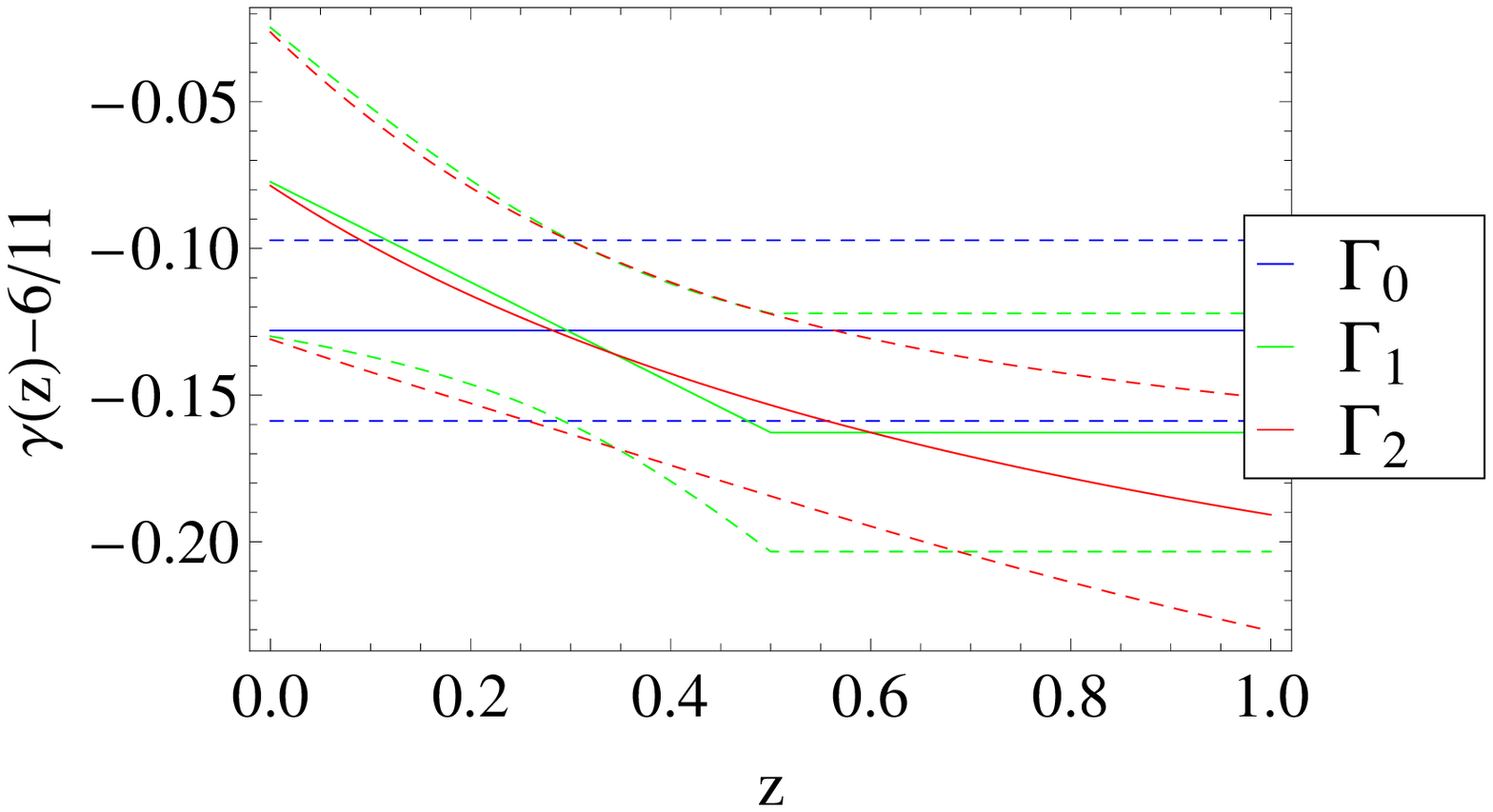}}}
\caption{The evolution of the growth index $\gamma(z)-\frac{6}{11}$ for the $f_{1-4}$CDM models [$f_{1}$CDM (top left), $f_{2}$CDM (top right), $f_{3}$CDM (bottom left), $f_{4}$CDM (bottom right)] and the various growth rate parameterizations. The lines correspond to $\Gamma_{0}$ (blue), $\Gamma_{1}$ (green), and $\Gamma_{2}$ (red). \label{fig:growth}}
\end{figure*}

\begin{figure*}[!]
\centering
\vspace{0cm}\rotatebox{0}{\vspace{0cm}\hspace{0cm}\resizebox{1\textwidth}{!}{\includegraphics{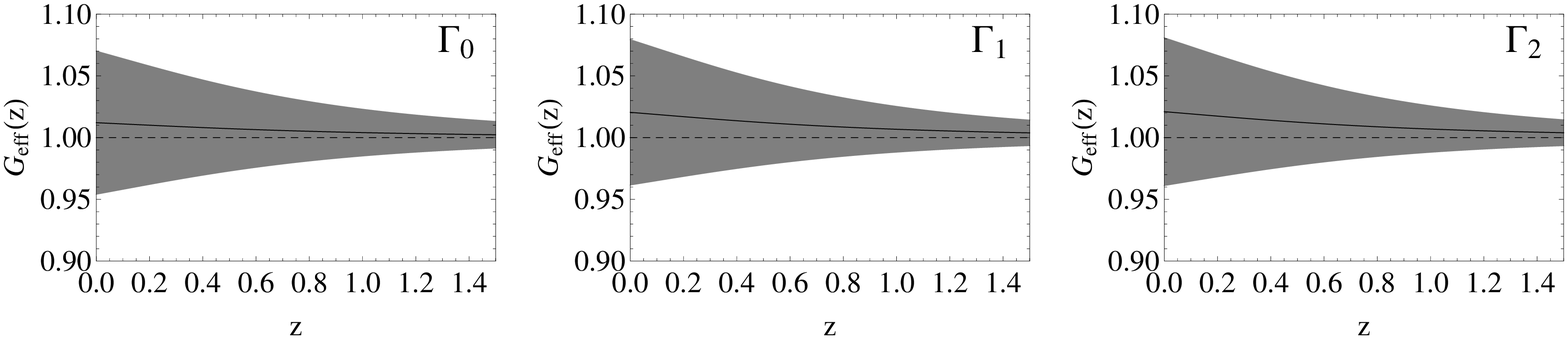}}}
\vspace{0cm}\rotatebox{0}{\vspace{0cm}\hspace{0cm}\resizebox{1\textwidth}{!}{\includegraphics{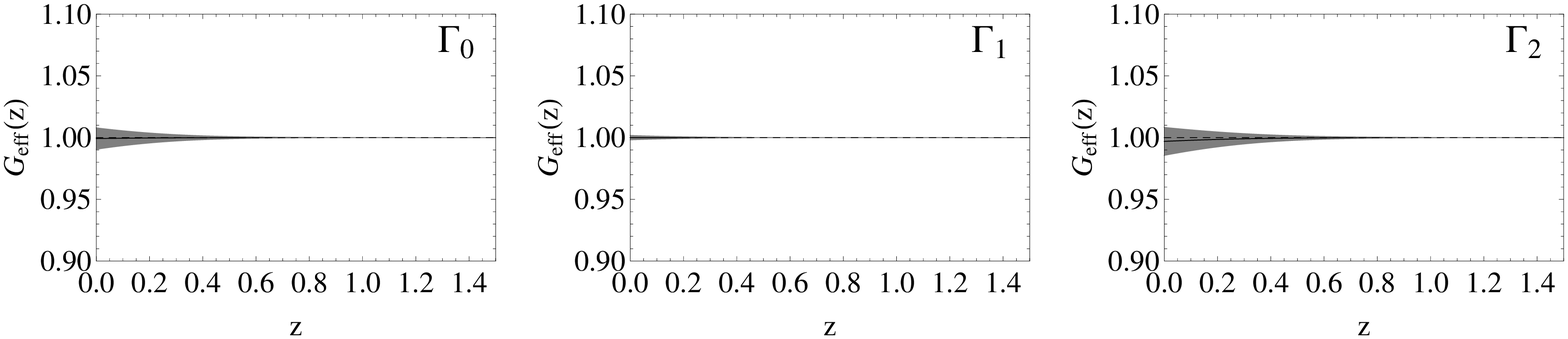}}}
\vspace{0cm}\rotatebox{0}{\vspace{0cm}\hspace{0cm}\resizebox{1\textwidth}{!}{\includegraphics{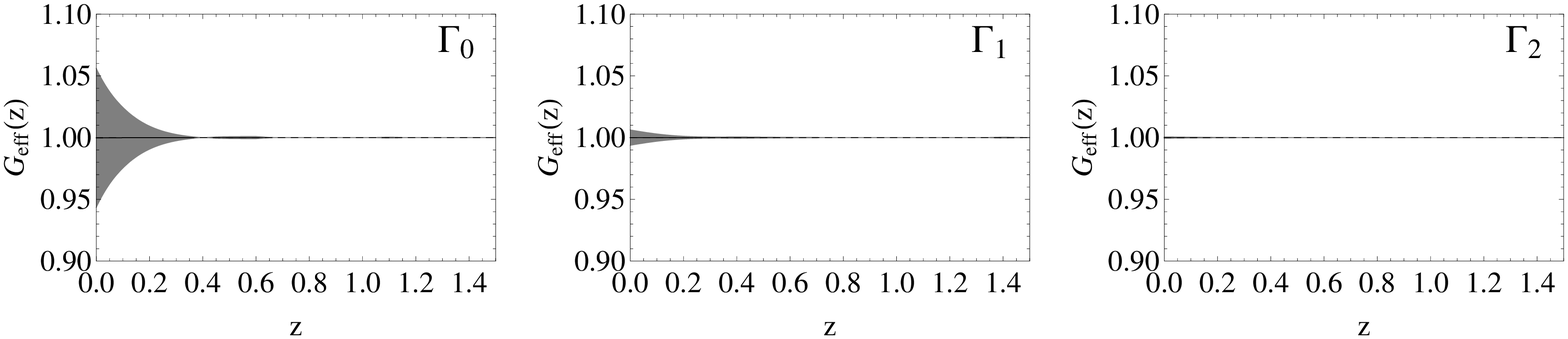}}}
\caption{The evolution of the $G_{\rm eff}(z)$ for the $f_{1-3}$CDM models and the various growth rate parameterizations considered in the text, $f_{1}$CDM (top), $f_{2}$CDM (middle), $f_{3}$CDM (bottom), for all three growth rate parametrizations $\Gamma_{0}$ (left), $\Gamma_{1}$ (middle), and
$\Gamma_{2}$ (right). The remarkable agreement between $G_{\rm eff}(z)$ and unity for the $f_{2}$CDM and  $f_{3}$CDM models is easily explained by the fact that these models exhibit little deviation from $\Lambda$CDM, as is easily seen in Table \ref{tab:growth1}.\label{fig:geff}}
\end{figure*}

For completeness, in Figs. \ref{growthrate}-\ref{fig:geff} we present a comparison of the observed and theoretical evolution of the growth rate $\fs(z)=F(z)\sigma_{8}(z)$, the evolution of the growth index $\gamma(z)-\frac{6}{11}$ and the evolution of the $G_{\rm eff}(z)$
respectively.


Finally, in order to enhance the validity of the above results, we repeat the
whole analysis by using $\sigma_8$ as a free parameter. As expected, we find
that the corresponding results are in good agreement, within $1\sigma$, with
those of $\sigma_8=0.8$ (see Table I). In particular, we find the following. \\

In the case of the $\Lambda$CDM,
\begin{itemize}
  \item for the $\Gamma_0$ model: $\chi^2=573.254$,  $\Omega_{m}= 0.272 \pm
0.003$, $\gamma_0=0.523 \pm 0.0858$, $\sigma_8= 0.761 \pm 0.038$;
  \item for the $\Gamma_1$ model: $\chi^2=572.618$,  $\Omega_{m}=
0.272\pm0.003$, $\gamma_0 = 0.485\pm0.098$, $\gamma_1 = -0.398\pm0.502$,
$\sigma_8 = 0.694\pm0.087$;
  \item for the $\Gamma_2$ model: $\chi^2=572.652$,  $\Omega_{m}= 0.272\pm
0.003$, $\gamma_0 = 0.483\pm0.097$, $\gamma_1 = -0.633\pm0.815$, $\sigma_8 =
0.685\pm0.097$;
\end{itemize}

In the case of the $f_1$CDM,
\begin{itemize}
  \item for the $\Gamma_0$ model: $\chi^2=573.618$,  $\Omega_{m}=
0.274\pm0.008$, $b = -0.019\pm0.087$, $\gamma_0 = 0.586\pm 0.090$, $\sigma_8
= 0.783\pm 0.041$;
  \item for the $\Gamma_1$ model: $\chi^2=576.124$,  $\Omega_{m}=
0.281\pm0.009$, $b = -0.099\pm0.109$, $\gamma_0 = 0.582\pm 0.092$, $\gamma_1
= 0.680\pm 0.443$, $\sigma_8 = 0.752\pm 0.070$;
  \item for the $\Gamma_2$ model: $\chi^2=573.756$,  $\Omega_{m}=
0.281\pm0.008$, $b = -0.098\pm0.104$, $\gamma_0 = 0.569\pm 0.103$, $\gamma_1
= 0.077\pm 0.872$, $\sigma_8 = 0.774\pm 0.114$;
\end{itemize}

In the case of the $f_2$CDM,
\begin{itemize}
  \item for the $\Gamma_0$ model: $\chi^2=573.264$,  $\Omega_{m}=
0.272\pm0.003$, $b = 0.101\pm0.186$,  $\gamma_0 = 0.523\pm0.086$, $\sigma_8 =
0.762\pm0.038$;
  \item for the $\Gamma_1$ model: $\chi^2=572.618$,  $\Omega_{m}=
0.272\pm0.003$, $b = 0.052\pm2.833$,  $\gamma_0 = 0.485\pm0.098$, $\gamma_1 =
-0.398\pm0.502$, $\sigma_8 = 0.694\pm 0.087$;
  \item for the $\Gamma_2$ model: $\chi^2=572.817$,  $\Omega_{m}=
0.272\pm0.003$, $b = 0.040\pm10.476$, $\gamma_0 = 0.500\pm0.113$, $\gamma_1 =
-0.599\pm1.022$, $\sigma_8 = 0.699\pm 0.127$;
\end{itemize}

In the case of the $f_3$CDM,
\begin{itemize}
  \item for the $\Gamma_0$ model: $\chi^2=573.224$,  $\Omega_{m}=
0.273\pm0.003$, $b = 0.050\pm2.561$, $\gamma_0 = 0.523\pm0.086$, $\sigma_8 =
0.761\pm0.038$;
  \item for the $\Gamma_1$ model: $\chi^2=572.599$,  $\Omega_{m}=
0.273\pm0.003$, $b = 0.051\pm2.264$, $\gamma_0 = 0.485\pm0.098$, $\gamma_1 =
-0.398\pm0.502$, $\sigma_8 = 0.694\pm 0.087$;
  \item for the $\Gamma_2$ model: $\chi^2=572.636$,  $\Omega_{m}=
0.273\pm0.003$, $b = 0.039\pm4.180$, $\gamma_0 = 0.486\pm0.098$, $\gamma_1 =
-0.598\pm0.817$, $\sigma_8 = 0.688\pm 0.098$;
\end{itemize}

In the case of the $f_4$CDM,
\begin{itemize}
  \item for the $\Gamma_0$ model: $\chi^2=703.539$, $\Omega_{m}= 0.202\pm0.002$, $\gamma_0 = 0.490\pm0.083$, $\sigma_8 = 0.856\pm0.061$;
  \item for the $\Gamma_1$ model: $\chi^2=702.419$, $\Omega_{m}= 0.202\pm0.002$, $\gamma_0 = 0.399\pm0.113$, $\gamma_1 = -0.418\pm0.401$, $\sigma_8 = 0.703\pm 0.134$;
  \item for the $\Gamma_2$ model: $\chi^2=702.501$, $\Omega_{m}= 0.202\pm0.002$, $\gamma_0 = 0.379\pm0.123$, $\gamma_1 = -0.733\pm0.713$, $\sigma_8 = 0.667\pm 0.154$;
\end{itemize}

In the case of the $f_5$CDM,
\begin{itemize}
  \item for the $\Gamma_0$ model: $\chi^2=577.279$, $\Omega_{m}= 0.285\pm0.006$, $b = 0.217\pm0.067$, $\gamma_0 = 0.550\pm0.086$, $\sigma_8 = 0.765\pm0.038$;
  \item for the $\Gamma_1$ model: $\chi^2=577.176$, $\Omega_{m}= 0.287\pm0.007$, $b = 0.189\pm0.076$, $\gamma_0 = 0.524\pm0.092$, $\gamma_1 = 0.057\pm0.470$, $\sigma_8 = 0.758\pm 0.083$;
  \item for the $\Gamma_2$ model: $\chi^2=575.983$, $\Omega_{m}= 0.287\pm0.007$, $b = 0.189\pm0.076$, $\gamma_0 = 0.489\pm0.090$, $\gamma_1 = -0.717\pm0.743$, $\sigma_8 = 0.674\pm 0.078$.
\end{itemize}

Lastly, we would like to emphasize that in all cases explored here the value
of AIC$_{\Lambda}$($\sim 578.3$) is smaller than the corresponding one for
the various $f(T)$ models, which implies that the usual $\Lambda$CDM
cosmology ($\gamma_{\Lambda}=0.597$) seems to provide a better fit than the
$f_{1-3}$CDM gravity models the expansion and the growth data. On the other hand,
the $|\Delta {\rm AIC}|$=$|{\rm AIC}_{\Lambda}-{\rm AIC}_{f_{1-3}(T)}|$ values
point that the growth data can be consistent with the $f_{1-3}$CDM gravity models.
We stress here that the  $f_4$CDM and $f_5$CDM  models seem to be
disfavored by the current data.

\section{Discussion and Conclusions}
\label{conclusions}

We have investigated  a wide range of different $f(T)$ models, with up to two
parameters, both at the background and at the perturbation level.
The functional forms of $f(T)$ considered in this work cover
practically all the functional forms considered in the literature so
far. Despite the fact that the $f(T)$ gravity can be derived from the principle
of least action the corresponding $f(T)$ functional forms
are phenomenological and even though they do not
correspond to a firm theoretical model they cover a wide range of
independent functional forms. Thus they represent a wide range of
degrees of freedom describing deviations from $\Lambda$CDM in the context of
$f(T)$ models.

Following our previous work Basilakos, Nesseris and Perivolaropoulos \cite{BasNes13} corresponding to $f(R)$ gravity, we calculated the function $y(z,b)$ which
quantifies the deviation from $\Lambda$CDM cosmology at the background level.
We also obtained the growth index and the effective Newton constant, which
incorporate the $f(T)$ gravity effects at the perturbation level. Furthermore,
we utilized the recent expansion and growth data, implementing the Akaike
information criterion and three different parametrizations for the growth
index, in order to constraint the parameters of these $f(T)$ models.

Our results show that all viable $f(T)$ gravity models hardly deviate from the $\Lambda$CDM paradigm. In particular, among the five
examined models, the power-law one \cite{Ben09} ($f_1$CDM), the
exponential-square-root one \cite{Linder:2010py} ($f_2$CDM) and the
exponential one ($f_3$CDM) possess $\Lambda$CDM cosmology as a limiting case. It is only this limit that is favored by cosmological observations.
In fact, the detailed observational
confrontation showed that these three models at best fit, behave as  small
perturbations around the concordance \lcdm cosmology, with the parameter $b$,
which quantifies the deviation from $\Lambda$CDM, constrained in
a very narrow window around $0$. The other two $f(T)$
models, namely the  logarithmic one \cite{Bamba}
($f_{4}$CDM) and the hyperbolic-tangent one
\cite{Wu:2011} ($f_{5}$CDM), do not possess $\Lambda$CDM
as a limiting case. We showed that both are in tension with the data.
In fact, we have demonstrated that ($f_{4}$CDM) coincides with the DGP model at the
background level, whose inconsistency  between distance measures
and horizon scale growth is well known \cite{Fang:2008kc} and
also demonstrated by our results.

The derived requirement of fine-tuning of the $f(T)$ constructions at the
$\Lambda$CDM, based on cosmological constraints, would probably be further
amplified if we had considered in addition their consistency with Solar
System tests, which constitute another powerful source of constraints against
any deviation from general relativity. At this point we would like to make a comment
concerning the Lorentz invariance of $f(T)$ theories.
As was shown in \cite{Li:2010cg}, for general $f(T)$ modifications the
field equations are not invariant under local Lorentz transformations,
unless $f(T)$ is a constant or a linear-in-$T$ function, in which case we
reobtain general relativity (that is, $\Lambda$CDM) and local Lorentz
invariance is restored. This feature imposes strict constraints on the
viable $f(T)$ forms, since the observational bounds on gravitational
Lorentz violation are very narrow \cite{Will:2005va}. As we have already
mentioned above, confrontation with Solar System data implies that the nontrivial $f(T)$
modification must be significantly small \cite{Iorio:2012cm}. In the
present analysis we were interested in performing a pure confrontation of
$f(T)$ theories with cosmological data, without imposing any other
theoretical constraints. Thus, from another point of view we verified again that
in all viable $f(T)$ scenarios the nontrivial $f(T)$ modifications are so
small that these constructions are practically indistinguishable from
$\Lambda$CDM. Clearly, taking into account the above Lorentz violation
discussion strengthens our result that all viable $f(T)$ almost coincide
with $\Lambda$CDM.

It is therefore safe to conclude that although at early times the
additional degrees of freedom provided by $f(T)$ constructions may play
an important role and improve the inflationary behavior, at late times
these extra degrees of freedom do not appear to be consistent with the
degrees of freedom favored by nature.

\section*{Acknowledgements}

The authors would like to thank Q.-G.Huang and C.-C. Lee for useful comments.
S.B. acknowledges support by the Research Center for Astronomy of the Academy
of Athens in the context of the program ``{\it Tracing the Cosmic
Acceleration}''. S.N. acknowledges financial support from the Madrid Regional
Government (CAM) under the program HEPHACOS S2009/ESP-1473-02, from MICINN
under Grant No. AYA2009-13936-C06-06 and Consolider-Ingenio 2010 PAU
(CSD2007-00060), as well as from the European Union Marie Curie Initial
Training Network UNILHC PITN-GA-2009-237920. S.N. also acknowledges the support
of the Spanish MINECO's ``Centro de Excelencia Severo Ochoa" Programme under
Grant No. SEV-2012-0249. The research of E.N.S. is implemented within the framework of the Action ``Supporting Postdoctoral Researchers'' of the Operational Program ``Education and Lifelong Learning'' Actionâs Beneficiary: (General Secretariat for Research and Technology), and is cofinanced by the European Social Fund (ESF) and the Greek State. This research has been cofinanced by the European Union (European Social Fund - ESF) and Greek national funds through the Operational Program "Education and Lifelong Learning" of the National Strategic Reference Framework (NSRF) - Research Funding Program: THALIS.\@ Investing in the society of knowledge through the European Social Fund. \\

\appendix
\section{DERIVATION OF EQ.~(\ref{approxM2})}
We can rewrite Eq.~(\ref{modfriedf1}) as
\bea
E^2(z)&=&\Omega_{m0}(1+z)^3+\Omega_{r0}(1+z)^4+\Omega_{F0} E^{2b}(z) \nn \\
&=&\Omega_{m0}(1+z)^3+\Omega_{r0}(1+z)^4+\Omega_{F0}-\Omega_{F0}\nn\\&+&\Omega_{F0} E^{2b}(z) \nn \\
&=&E^2_\Lambda(z)+\Omega_{F0}\left[E^{2b}(z)-1\right], \label{friedm11}\eea
where $E^2_\Lambda(z)$ is given by Eq.~(\ref{friedlcdm}) and in the second line we added and subtracted $\Omega_{F0}$.

Now, in this case we assume that the Hubble parameter
$\frac{H^2}{H_0^2}\equiv E^2(z)$ depends on $b$ only implicitly via the
Friedmann equation  (\ref{friedlcdm}). In other words, we consider $b$ and
$E^2(z)$ to be independent, and thus any derivatives with respect to
$b$ are   zero. Hence, performing a Taylor expansion of
 (\ref{friedm11}) up to second order around $b=0$ we acquire
\bea
E^2(z)&=&E^2_\Lambda(z)+\ln\left[E^2(z)\right]\Omega_{F0}~b\nn\\&+&\frac{1}{2
} \ln\left[E^2(z)\right]^2\Omega_{F0}~b^2+\cdots .
\eea
If we keep only the first-order term and solve for $E^2(z)$, we obtain
\be
E^2(z,b)=-b~\Omega_{F0}~\mathcal{W}_k\left(-\frac{e^{-\frac{E_{\Lambda
}(z){}^2}{b~\Omega_{F0}}}}{b\;\Omega_{F0}}\right),
\ee
where $\mathcal{W}_k(\omega)$ is the Lambert function defined via
$\omega\equiv\mathcal{W}_k(\omega)e^{\mathcal{W}_k(\omega)}$ for all
complex numbers $\omega$. The Lambert function has branch-cut discontinuities, so the different branches are indicated by the integer $k$. Our solution has $k=0$ (the principal branch) for $b\leq0$ and $k=-1$ for $b>0$.


\begin{thebibliography}{99}

\bibitem{Capozziello:2011et}
  S.~Capozziello and M.~De Laurentis,
  Phys.\ Rept.\  {\bf 509}, 167 (2011).

\bibitem{Ame10}E. J. Copeland, M. Sami and S. Tsujikawa,
Intern. Journal of
Modern Physics D, \textbf{15}, 1753,(2006);
R. R. Caldwell and M. Kamionkowski, Ann.Rev.Nucl.Part.Sci.,
\textbf{59}, 397 (2009);
I. Sawicki and W. Hu, Phys. Rev. D, {\bf 75}, 127502 (2007).


\bibitem{Ame10b}
L. Amendola and S. Tsujikawa,
{\it{Dark
Energy Theory and Observations}}, Cambridge University Press, Cambridge UK,
(2010).

\bibitem{ein28}
A. Einstein, Sitz. Preuss. Akad. Wiss. p. \textbf{17}, 217 (1928); \textbf{17} 224 (1928);
  A.~Unzicker and T.~Case,
  physics/0503046.

\bibitem{Hayashi79}
  K. Hayashi and T. Shirafuji,
  Phys. Rev. D \textbf{19}, 3524 (1979);
  Addendum-ibid. \textbf{24}, 3312 (1981).

\bibitem{Maluf:1994ji}
  J.~W.~Maluf,
  J.\ Math.\ Phys.\ \textbf{35} (1994) 335;
  H.~I.~Arcos and J.~G.~Pereira,
  Int.\ J.\ Mod.\ Phys.\ D \textbf{13}, 2193 (2004).

\bibitem{Ferraro:2006jd}
  R.~Ferraro and F.~Fiorini,
  Phys.\ Rev.\ D {\bf 75}, 084031 (2007);
  R.~Ferraro, F.~Fiorini,
  Phys.\ Rev.\  {\bf D78}, 124019 (2008).

\bibitem{Ben09}
G. R. Bengochea, \& R. Ferraro, Phys. Rev. D, {\bf 79}, 124019, (2009).

\bibitem{Linder:2010py}
  E.~V.~Linder,
  Phys.\ Rev.\ D \textbf{81}, 127301 (2010); Erratum,
Phys. Rev. D, {\bf 82}, 109902.

\bibitem{Myrzakulov:2010vz}
  K.~K.~Yerzhanov, S.~.R.~Myrzakul, I.~I.~Kulnazarov and R.~Myrzakulov,
  arXiv:1006.3879 [gr-qc];
  K.~Bamba, C.~-Q.~Geng and C.~-C.~Lee,
  arXiv:1008.4036 [astro-ph.CO];
  R.~-J.~Yang,
  Europhys.\ Lett.\ \textbf{93}, 60001 (2011);
   Y.~Zhang, H.~Li, Y.~Gong, Z.~-H.~Zhu,
  JCAP {\bf 1107}, 015 (2011);
   R.~Ferraro, F.~Fiorini,
  Phys.\ Lett.\  {\bf B702}, 75 (2011).
  Y.~-F.~Cai, S.~-H.~Chen, J.~B.~Dent, S.~Dutta, E.~N.~Saridakis,
  Class.\ Quant.\ Grav.\  {\bf 28}, 2150011 (2011);
  M.~Sharif, S.~Rani,
  Mod.\ Phys.\ Lett.\  {\bf A26}, 1657 (2011);
   S.~Capozziello, V.~F.~Cardone, H.~Farajollahi and A.~Ravanpak,
  Phys.\ Rev.\ D {\bf 84}, 043527 (2011);
  K.~Bamba and C.~-Q.~Geng,
  JCAP {\bf 1111}, 008 (2011);
  C.~-Q.~Geng, C.~-C.~Lee, E.~N.~Saridakis, Y.~-P.~Wu,
  Phys.\ Lett.\  {\bf B704}, 384 (2011);
    H.~Wei,
  Phys.\ Lett.\ B {\bf 712}, 430 (2012);
  C.~-Q.~Geng, C.~-C.~Lee, E.~N.~Saridakis,
  JCAP {\bf 1201}, 002 (2012);
    Y.~-P.~Wu and C.~-Q.~Geng,
  Phys.\ Rev.\ D {\bf 86}, 104058 (2012);
  C.~G.~Bohmer, T.~Harko and F.~S.~N.~Lobo,
  Phys.\ Rev.\ D {\bf 85}, 044033 (2012);
  H.~Farajollahi, A.~Ravanpak and P.~Wu,
  Astrophys.\ Space Sci.\  {\bf 338}, 23 (2012);
  K.~Atazadeh and F.~Darabi,
  Eur.\ Phys.\ J.\ C {\bf 72}, 2016 (2012);
   M.~Jamil, D.~Momeni, N.~S.~Serikbayev and R.~Myrzakulov,
  Astrophys.\ Space Sci.\  {\bf 339}, 37 (2012);
  J.~Yang, Y.~-L.~Li, Y.~Zhong and Y.~Li,
  arXiv:1202.0129 [hep-th];
    K.~Karami and A.~Abdolmaleki,
  JCAP {\bf 1204}, 007 (2012);
   C.~Xu, E.~N.~Saridakis and G.~Leon,
  JCAP {\bf 1207}, 005 (2012);
  K.~Bamba, R.~Myrzakulov, S.~'i.~Nojiri and S.~D.~Odintsov,
  arXiv:1202.4057 [physics.gen-ph];
   D.~Liu, P.~Wu and H.~Yu,
  Int.\ J.\ Mod.\ Phys.\ D {\bf 21}, 1250074 (2012);
  H.~Dong, Y.~-b.~Wang and X.~-h.~Meng,
  Eur.\ Phys.\ J.\ C {\bf 72}, 2002 (2012);
  N.~Tamanini and C.~G.~Boehmer,
  Phys.\ Rev.\ D {\bf 86}, 044009 (2012);
  K.~Bamba, S.~Capozziello, S.~'i.~Nojiri and S.~D.~Odintsov,
  Astrophys.\ Space Sci.\  {\bf 342}, 155 (2012);
  A.~Behboodi, S.~Akhshabi and K.~Nozari,
  Phys.\ Lett.\ B {\bf 718}, 30 (2012);
  A.~Banijamali and B.~Fazlpour,
  Astrophys.\ Space Sci.\  {\bf 342}, 229 (2012);
  D.~Liu and M.~J.~Reboucas,
  Phys.\ Rev.\ D {\bf 86}, 083515 (2012);
  M.~E.~Rodrigues, M.~J.~S.~Houndjo, D.~Saez-Gomez and F.~Rahaman,
  Phys.\ Rev.\ D {\bf 86}, 104059 (2012);
  Y.~-P.~Wu and C.~-Q.~Geng,
  arXiv:1211.1778 [gr-qc];
      S.~Chattopadhyay and A.~Pasqua,
 Astrophys.\ Space Sci.\  {\bf 344}, 269 (2013);
  M.~Jamil, D.~Momeni and R.~Myrzakulov,
  Gen.\ Rel.\ Grav.\  {\bf 45}, 263 (2013);
  K.~Bamba, J.~de Haro and S.~D.~Odintsov,
  JCAP {\bf 1302}, 008 (2013);
  M.~Jamil, D.~Momeni and R.~Myrzakulov,
  Eur.\ Phys.\ J.\ C {\bf 72}, 2267 (2012);
  J.~-T.~Li, C.~-C.~Lee and C.~-Q.~Geng,
  Eur.\ Phys.\ J.\ C {\bf 73}, 2315 (2013);
   H.~M.~Sadjadi,
 Phys.\ Rev.\ D {\bf 87}, 064028 (2013);
  A.~Aviles, A.~Bravetti, S.~Capozziello and O.~Luongo,
  Phys.\ Rev.\ D {\bf 87}, 064025 (2013);
   Y.~C.~Ong, K.~Izumi, J.~M.~Nester and P.~Chen,
 Phys.\ Rev.\ D {\bf 88}, 024019 (2013);
  K.~Bamba, S.~'i.~Nojiri and S.~D.~Odintsov,
  arXiv:1304.6191 [gr-qc];
      H.~Dong, J.~Wang and X.~Meng,
 arXiv:1304.6587 [gr-qc];
 G.~Otalora,
 JCAP {\bf 1307}, 044 (2013);
  J.~Amoros, J.~de Haro and S.~D.~Odintsov,
     Phys.\ Rev.\ D {\bf 87}, 104037 (2013);
    F.~Darabi,
 arXiv:1305.5378 [gr-qc];
  G.~Otalora,
  arXiv:1305.5896 [gr-qc];
  C.~-Q.~Geng, J.~-A.~Gu and C.~-C.~Lee,
   Phys.\ Rev.\ D {\bf 88}, 024030 (2013);
  I.~G.~Salako, M.~E.~Rodrigues, A.~V.~Kpadonou, M.~J.~S.~Houndjo and
J.~Tossa,
 arXiv:1307.0730 [gr-qc];
   A.~V.~Astashenok,
 arXiv:1308.0581 [gr-qc];
  M.~E.~Rodrigues, I.~G.~Salako, M.~J.~S.~Houndjo and J.~Tossa,
 arXiv:1308.2962 [gr-qc].


\bibitem{Wu:2010mn}
  P.~Wu, H.~W.~Yu,
  Phys.\ Lett.\ \textbf{B693}, 415 (2010).



\bibitem{Bengochea001}
    G.~R.~Bengochea,
  Phys.\ Lett.\  {\bf B695}, 405 (2011).

\bibitem{Iorio:2012cm}
  L.~Iorio and E.~N.~Saridakis,
  Mon.\  Not.\  Roy.\  Astron.\  Soc.\  {\bf 427}, 1555 (2012).


\bibitem{Wang:2011xf}
    T.~Wang,
  Phys.\ Rev.\  {\bf D84}, 024042 (2011);
  R.~-X.~Miao, M.~Li and Y.~-G.~Miao,
  JCAP {\bf 1111}, 033 (2011);
  C.~G.~Boehmer, A.~Mussa and N.~Tamanini,
  Class.\ Quant.\ Grav.\  {\bf 28}, 245020 (2011);
  M.~Hamani Daouda, M.~E.~Rodrigues and M.~J.~S.~Houndjo,
  Eur.\ Phys.\ J.\ C {\bf 71}, 1817 (2011);
  R.~Ferraro, F.~Fiorini,
  Phys.\ Rev.\ D {\bf 84}, 083518 (2011);
  M.~H.~Daouda, M.~E.~Rodrigues and M.~J.~S.~Houndjo,
  Eur.\ Phys.\ J.\ C {\bf 72}, 1890 (2012);
  P.~A.~Gonzalez, E.~N.~Saridakis and Y.~Vasquez,
  JHEP {\bf 1207}, 053 (2012);
  H.~Wei, X.~-J.~Guo and L.~-F.~Wang,
  Phys.\ Lett.\ B {\bf 707}, 298 (2012);
  S.~Capozziello, P.~A.~Gonzalez, E.~N.~Saridakis and Y.~Vasquez,
  JHEP {\bf 1302} (2013) 039;
  K.~Atazadeh and M.~Mousavi,
  Eur.\ Phys.\ J.\ C {\bf 72}, 2272 (2012).


\bibitem{Zhang:2012jsa}
  W.~-S.~Zhang, C.~Cheng, Q.~-G.~Huang, M.~Li, S.~Li, X.~-D.~Li and S.~Wang,
  Sci.\ China Phys.\ Mech.\ Astron.\  {\bf 55}, 2244 (2012).

\bibitem{Cardone:2012xq}
  V.~F.~Cardone, N.~Radicella and S.~Camera,
  Phys.\ Rev.\ D {\bf 85}, 124007 (2012).

\bibitem{Starobinsky:2007hu}
  A.~A.~Starobinsky,
  JETP Lett.\  {\bf 86}, 157 (2007).


\bibitem{BasNes13}
  S.~Basilakos, S.~Nesseris and L.~Perivolaropoulos,
Phys.\ Rev.\ D {\bf 87}, 123529 (2013).

\bibitem {Hu07}
W. Hu and I. Sawicki, Phys. Rev. D, \textbf{76}, 064004 (2007).

\bibitem{Weitzenb23}
  Weitzenb\"{o}ck R.,
  \emph{Invarianten Theorie},
  Nordhoff, Groningen (1923).

\bibitem{Dave02}
R. Dave, R. R. Caldwell and P. J. Steinhardt, Phys. Rev. D, {\bf 66}, 023516
(2002).


\bibitem{Gann09}
R. Gannouji, B. Moraes and D. Polarski, JCAP, {\bf 62}, 034 (2009).

\bibitem{Lue04}
A. Lue, R. Scossimarro, and G. D. Starkman, Phys. Rev. D, {\bf 69}, 124015
(2004).

\bibitem{Linder05}
E. V. Linder, Phys. Rev. D, {\bf 72}, 043529 (2005).

\bibitem {Stab06}
F. H. Stabenau  and B. Jain, Phys. Rev. D, {\bf 74}, 084007 (2006).

\bibitem {Uzan07}
P. J. Uzan, Gen.\ Rel.\ Grav., {\bf 39}, 307 (2007).

\bibitem {Tsu08}
S. Tsujikawa, K. Uddin and R. Tavakol, Phys. Rev. D, {\bf 77}, 043007
(2008).

\bibitem{Peeb93}
P. J. E. Peebles, {\it{Principles of Physical Cosmology}}, Princeton
University Press, Princeton New Jersey (1993).

\bibitem{Silv94}
V. Silveira and I. Waga, Phys. Rev. D, {\bf 50}, 4890 (1994).

\bibitem{Wang98}
L.~Wang and J.~P.~Steinhardt, Astrophys.\ J,\ {\bf 508}, 483 (1998).

\bibitem{Linder2007}
E. V. Linder, Phys. Rev. D, {\bf 70}, 023511 (2004);
E. V. Linder, and R. N. Cahn, Astrop. \ Phys., {\bf 28}, 481 (2007).

\bibitem{Nes08}
  S.~Nesseris and  L.~Perivolaropoulos,
  Phys.\ Rev.\ D {\bf 77}, 023504 (2008).



  \bibitem{Dvali2000}
G. Dvali, G. Gabadadze, M. Porrati, Phys. Lett. B {\bf
485}, 208 (2000).


\bibitem{Gong10}
Y. Gong, Phys. Rev. D, {\bf 78} 123010 (2008).

\bibitem{Wei08}
H. Wei, Phys. Lett. B., {\bf 664}, 1 (2008).


\bibitem{Fu09}
Y.G. Gong, Phys. Rev. D, {\bf 78}, 123010 (2008)
X.-y Fu, P.-x Wu and H.-w, Phys. Lett. B., {\bf 677}, 12 (2009).

\bibitem{Tsu09}
S. Tsujikawa, R. Gannouji, B. Moraes and D. Polarski,
Phys. Rev. D, {\bf 80}, 084044 (2009).


\bibitem{Bastav13}
S. Basilakos and P. Stavrinos,
Phys. Rev. D, {\bf 87}, 043506 (2013).

\bibitem{Saini00}
T. D. Saini, S. Raychaudhury,V. Sahni, and A. A. Starobinsky,
Phys. Rev. Lett., {\bf 85}, 1162, (2000);
D. Huterer, and M. S. Turner, Phys. Rev. D, {\bf 64}, 123527 (2001).

\bibitem{Pol}
D.~Polarski and R.~Gannouji, Phys.\ Lett.\ B {\bf 660}, 439 (2008).

\bibitem{Bel12}
A. B. Belloso, J. Garcia-Bellido and D. Sapone, JCAP, {\bf 1110}, 010 (2011).

\bibitem{DP11}
C. Di Porto, L. Amendola and E. Branchini, Mon.\ Not.\ Roy.\
Astron.\ Soc.\ {\bf 419}, 985 (2012).


\bibitem{Ishak09}
M. Ishak and J. Dosset, Phys. Rev. D, {\bf 80}, 043004 (2009).



\bibitem{Zheng:2010am}
  R.~Zheng, Q.~-G.~Huang,
  JCAP \textbf{1103}, 002 (2011).


\bibitem{Myrzakulov}
R. Myrzakulov, Entropy, {\bf 14}, 1627 (2012), arXiv:1212.2155

\bibitem{Ferraro}
R. Ferraro, AIP Conf. Proc. {\bf 1471}, 103 (2012), arXiv:1204.6273


\bibitem{Chen001}
S.~H.~Chen, J.~B.~Dent, S.~Dutta and E.~N.~Saridakis,
Phys.\ Rev.\ D
\textbf{ 83}, 023508 (2011);
J.~B.~Dent, S.~Dutta, E.~N.~Saridakis,
  JCAP {\bf 1101}, 009 (2011).
  K.~Izumi and Y.~C.~Ong,
 JCAP {\bf 1306}, 029 (2013).

  \bibitem{Linn1}
E. V. Linder, Phys. Rev. D, {\bf 80}, 123528, (2009).



\bibitem{Bamba}
Bamba, K., Geng Chao-Qiang, Lee Chung-Chi, Luo Ling-Wei,
  JCAP \textbf{1101}, 021
 (2011).


\bibitem{Fang:2008kc}
  W.~Fang, S.~Wang, W.~Hu, Z.~Haiman, L.~Hui and M.~May,
  Phys.\ Rev.\ D {\bf 78}, 103509 (2008).


  \bibitem{Wu:2011}
P.~Wu and H.~W.~Yu,
Eur.\ Phys.\ J.\ C {\bf 71}, 1552 (2011).



\bibitem{Akaike1974}
H. Akaike, IEEE Transactions of Automatic Control,
  {\bf 19}, 716 (1974); N. Sugiura, Communications in Statistics A, Theory
and Methods, {\bf 7}, 13 (1978).


\bibitem{Suzuki:2011hu}
  N.~Suzuki, D.~Rubin, C.~Lidman, G.~Aldering, R.~Amanullah,  K.~Barbary,
L.~F.~Barrientos and J.~Botyanszki {\it et al.},
  Astrophys.\ J\  {\bf 746}, 85 (2012).

\bibitem{Blake:2011en}
  C.~Blake, E.~Kazin, F.~Beutler, T.~Davis, D.~Parkinson, S.~Brough,
M.~Colless and C.~Contreras {\it et al.},
  Mon.\ Not.\ Roy.\ Astron.\ Soc.\  {\bf 418}, 1707 (2011).

\bibitem{Perc10} W. J. Percival, Mon. Not. Roy. Astron. Soc., {\bf 401}
2148 (2010).

\bibitem{Hinshaw:2012fq}
G.~Hinshaw {\it et al.}  [WMAP Collaboration],
  Astrophys.\ J.\ Suppl.\  {\bf 208}, 19 (2013)
  [arXiv:1212.5226 [astro-ph.CO]].

\bibitem{Sam11}
  L.~Samushia, W.~J.~Percival and A.~Raccanelli,
  Mon.\ Not.\ Roy.\ Astron.\ Soc.\  {\bf 420}, 2102 (2012).

\bibitem{Bass}
  S.~Basilakos,
  Int.\ J.\ Mod.\ Phys.\ D {\bf 21}, 1250064 (2012).


\bibitem{Por}
S. Basilakos and A. Pouri, Mon. Not. Roy. Astron. Soc., {\bf 423}, 3761
(2012).

\bibitem{Hud12}
M. J. Hudson and S. J. Turnbull, Astrophys. J. Let. {\bf 751}, 30 (2012).

\bibitem{Samnew12}
  L.~Samushia, B.~A.~Reid, M.~White, W.~J.~Percival, A.~J.~Cuesta,
L.~Lombriser, M.~Manera and R.~C.~Nichol {\it et al.},
  Mon.\ Not.\ Roy.\ Astron.\ Soc.\  {\bf 429}, 1514 (2013).

\bibitem{Port08}
C. Di Porto and L. Amendola,
  Phys. Rev. D, {\bf 77}, 083508  (2008).

\bibitem{Dos10}
  J.~Dossett, M.~Ishak, J.~Moldenhauer, Y.~Gong and
A.~Wang,
  JCAP {\bf 1004}, 022 (2010).


\bibitem{Li:2010cg}
  B.~Li, T.~P.~Sotiriou and J.~D.~Barrow,
Phys.\ Rev.\ D {\bf 83}, 064035 (2011);


\bibitem{Will:2005va}
  C.~M.~Will,
  Living Rev.\ Rel.\  {\bf 9}, 3 (2006);
  Q.~G.~Bailey, R.~D.~Everett and J.~M.~Overduin,
Phys.\ Rev.\ D \textbf{88}, 102001 (2013);


\end{thebibliography}
\end{document}